\documentclass[preprint]{aastex}

\shorttitle{Chandra Orion Ultradeep Project: Observations and
source lists}

\shortauthors{Getman et al.}

\slugcomment{Accepted for publication in the Astrophysical Journal
Supplement, the COUP Special Issue}

\begin{document}

\title{$Chandra$ Orion Ultradeep Project: Observations and Source
Lists}

\author{K.\ V.\ Getman\altaffilmark{1}, E.\ Flaccomio\altaffilmark{2},
P.\ S.\ Broos\altaffilmark{1}, N.\ Grosso\altaffilmark{3},  M.\
Tsujimoto\altaffilmark{1}, L.\ Townsley\altaffilmark{1},  G.\ P.\
Garmire\altaffilmark{1}, J.\ Kastner\altaffilmark{4}, J.\
Li\altaffilmark{4}, F.\ R.\ Harnden, Jr.\altaffilmark{5}, S.\
Wolk\altaffilmark{5}, S.\ S.\ Murray\altaffilmark{5}, C.\ J.\
Lada\altaffilmark{5}, A.\ A.\ Muench\altaffilmark{5}, M.\ J.\
McCaughrean\altaffilmark{6}, G.\ Meeus\altaffilmark{6}, F.\
Damiani\altaffilmark{2}, G.\ Micela\altaffilmark{2}, S.\
Sciortino\altaffilmark{2}, J.\ Bally\altaffilmark{7}, L.\ A.\
Hillenbrand\altaffilmark{8}, W.\ Herbst\altaffilmark{9}, T.\
Preibisch\altaffilmark{10}, E.\ D.\ Feigelson\altaffilmark{1}}

\altaffiltext{1}{Department of Astronomy \& Astrophysics,
Pennsylvania State University, 525 Davey Laboratory, University
Park, PA 16802} \altaffiltext{2}{INAF, Osservatorio Astronomico di
Palermo G. S. Vaiana, Piazza del Parlamento 1, I-90134 Palermo,
Italy} \altaffiltext{3}{Laboratoire d'Astrophysique de Grenoble,
Universite Joseph Fourier, BP 53, 38041 Grenoble Cedex 9, France}
\altaffiltext{4}{Chester F. Carlson Center for Imaging Science,
Rochester Institute of Technology, 54 Lomb Memorial Drive,
Rochester, NY 14623} \altaffiltext{5}{Smithsonian Astrophysical
Observatory, 60 Garden Street, Cambridge, MA 02138}
\altaffiltext{6}{Astrophysikalisches Institut Potsdam, An der
Sternwarte 16, D-14482 Potsdam, Germany} \altaffiltext{7}{Center
for Astrophysics and Space Astronomy, University of Colorado at
Boulder, CB 389, Boulder, CO 80309} \altaffiltext{8}{Department of
Astronomy, California Institute of Technology, Pasadena, CA
91125}\altaffiltext{9}{Max-Planck-Institut fur Astronomie,
Konigstuhl 17, 69117 Heidelberg, Germany}
\altaffiltext{10}{Max-Planck-Institut fur Radioastronomie, Auf dem
Hugel 69, 53121 Bonn, Germany}

\begin{abstract}
We present a description of the data reduction methods and the
derived catalog of more than 1600 X-ray point sources from the
exceptionally deep January 2003 Chandra X-ray Observatory
($Chandra$) observation of the Orion Nebula Cluster and embedded
populations around OMC-1. The observation was obtained with
$Chandra$'s Advanced CCD Imaging Spectrometer (ACIS) and has been
nicknamed the $Chandra$ Orion Ultradeep Project (COUP). With an
838~ks exposure made over a continuous period of 13.2 days, the
COUP observation provides the most uniform and comprehensive
dataset on the X-ray emission of normal stars ever obtained in the
history of X-ray astronomy.
\end{abstract}


\section{Introduction} \label{introduction_section}

The processes of star formation and early stellar evolution have
proved to be complex in many respects. Gravitational collapse can
be hindered by magnetic and turbulent pressures and is accompanied
by collimated jets and outflows during the protostellar phase. The
dynamics and evolution of circumstellar disks where planets form
are still under study.  The birth of stars rarely occurs in
isolation. Most stars form in a hierarchy ranging from binaries to
rich clusters. These processes, which take place in
thermodynamically cold and neutral media with characteristic
energies of $\ll1$ eV per particle, paradoxically produce and are
subject to violent high energy processes with characteristic
energies of $\gtrsim 10^3$~eV.  The principal evidence for this is
X-ray emission (and sometimes nonthermal radio emission) from
stars throughout their pre-main sequence (PMS) evolution
\citep{Feigelson99}.

PMS X-ray emission is elevated $10^1$ to $\gtrsim 10^4$ times
above typical main sequence levels for stars with masses $0.1
M_\odot \la M \la 2 M_\odot$. High-amplitude X-ray variability and
hard spectra, supported by multiwavelength studies such as
photospheric Zeeman measurements and starspot mapping, indicate
that the emission is primarily attributable to solar-type magnetic
flares where plasma is heated to high temperatures by violent
reconnection events in magnetic loops. However, the empirical
relationships between PMS X-ray emission and stellar age, mass,
radius, and rotation differ from those seen among solar-type main
sequence stars \citep{Flaccomio03b, Feigelson03}. We have an
inadequate understanding of the astrophysical origins of this
magnetic activity.  It is also unclear whether or not the high
energy radiation from magnetic flaring has important astrophysical
effects. For example, it is possible that some planets form in
disk regions that are rendered turbulent from MHD instabilities
induced by PMS X-rays \citep[e.g.,][]{Glassgold00, Matsumura03}.
X-rays from embedded stars may affect ambipolar diffusion, and
thus future star formation, in the surrounding molecular cloud.
X-ray images are also particularly effective in locating low mass
companions around luminous young stars. There are thus a variety
of astrophysical reasons to investigate the magnetic activity of
young stars.  Due to the wide range of activity levels and other
properties, it is advantageous to study large samples of stars.

The nearest rich and concentrated sample of PMS stars is the Orion
Nebula Cluster (ONC), also known as the Trapezium Cluster or Ori
Id OB association. The OB members of the ONC illuminate the Orion
Nebula (= Messier 42), a blister H~$_{\rm{II}}$ region at the near
edge of Orion A, the nearest giant molecular cloud ($D \simeq 450$
pc). The ONC has $\simeq 2000$ members within a 1 pc
($8^{\prime}$) radius sphere with 80\% of the stars younger than 1
Myr \citep{Hillenbrand97}. It was the first young stellar cluster
to be detected in the X-ray band \citep{Giacconi72} and
non-imaging studies soon found that the X-ray emission is extended
on scales of a parsec or larger \citep{denBoggende78,Bradt79}.
Early explanations for the Orion X-rays included winds from the
massive Trapezium stars colliding with each other or the molecular
cloud, and hot coronae or magnetic activity in lower mass T Tauri
stars.  The $Einstein$ \citep{Ku79}, $ROSAT$ \citep{Gagne95,
Geier95, Alcala96} and $ASCA$ \citep{Yamauchi96} imaging X-ray
observatories established that both the Trapezium stars and many
lower-mass T Tauri stars contribute to the X-ray emission.

But these early studies could detect only a modest fraction of the
ONC stars due to the high stellar densities and heavy absorption
by molecular material along the line of sight. These limitations
are alleviated by NASA's Chandra X-ray Observatory ($Chandra$).
Its mirrors provide an unprecedented combination of capabilities -
a wide spectral bandpass and $\lesssim 1$\arcsec\/ spatial
resolution - and are particularly well-suited to the study of
crowded and absorbed clusters of faint X-ray sources.  The ONC was
thus intensively studied during the first year of the $Chandra$
mission with several instrumental setups: the Advanced CCD Imaging
Spectrometer (ACIS) in imaging mode \citep{Garmire00,
Feigelson02a, Feigelson02b, Feigelson03} and as detector for the
High Energy Transmission Grating Spectrometer \citep{Schulz00,
Schulz01, Schulz03}, and with the High Resolution Imager
\citep[HRI;][]{Flaccomio03a, Flaccomio03b}.

While many valuable results emerged from these early $Chandra$
studies, it was recognized that more would be learned from a much
deeper and longer exposure of the Orion Nebula region.  We present
here a unique $\sim 10$ day, nearly continuous observation of the
Orion Nebula obtained during the fourth year of the mission,
nicknamed the $Chandra$ Orion Ultradeep Project (COUP). This paper
describes the COUP observations (\S \ref{observations_section}),
the data reduction (\S \ref{data_selection_section}), source
detection and characterization (\S
\ref{source_detection_section}-\ref{luminosities_section}), and
identification with previously known PMS stars (\S
\ref{stellar_counterparts_section}-\ref{nondetections_section}).
We provide extensive machine-readable tables of the X-ray sources
and properties, and an atlas of their location in the sky (\S
\ref{atlas_section}).  Other COUP papers will discuss various
astronomical and astrophysical issues emerging from these
observations.

\section{Observations}\label{observations_section}

The COUP combines six nearly-consecutive exposures of the Orion
Nebula Cluster (ONC) spanning 13.2 days in January 2003 with a
total exposure time of 838 ks or 9.7 days (Table
\ref{observation_journal_table}). The observations were obtained
with the ACIS camera on-board $Chandra$, which are described by
\citet{Weisskopf02} and \citet{Garmire03}. We consider here only
results arising from the imaging array (ACIS-I) of four abutted
$1024 \times 1024$ pixel front-side illuminated charge-coupled
devices (CCDs) covering about $17\arcmin \times 17 \arcmin$ on the
sky. The S2 chip in the spectroscopic array (ACIS-S) was also
operational, but as the telescope point spread function (PSF) is
considerably degraded far off-axis, the S2 data are omitted from
the analysis.  The ACIS-I aimpoint and the roll angle for all six
observations were kept constant (Table
\ref{observation_journal_table}). All six observations were taken
in Very Faint telemetry mode to improve the screening of
background events and thus increase the sensitivity of the
observations. The focal plane temperature was between $-119.5$ and
$-119.7$ $^\circ$C.

The observations were constrained to be performed in January to
accommodate a single roll of about 311 degrees.  The observational
span of 14 days means that the spacecraft's nominal roll changes
by about 14 degrees.  Typically, off-nominal rolls of greater than
$\pm 5^\circ$ are difficult to accommodate.  However in January of
2003 the tolerance was somewhat higher in the direction of the
ONC. The observations were continuous with the exception of five
passages through the Van Allen belts,  during each of which ACIS
is taken out of the focal plane for about 29 ks. The spacecraft
used these passages to unload momentum so that no other science
target was observed between Jan 8 2003 21:00 GMT and Jan 22 2002
02:00 GMT. The outcome of 9.7 days exposure in a 13.2 day period
(73\% efficiency) is the maximum efficiency achievable by the
spacecraft given the current perigee of about 33,000 km.

The COUP observation period was also remarkable in having some of
the quietest solar weather in the $Chandra$ mission.  This reduces
instrumental background and makes the dataset truly exceptional
for variability analysis. The Advanced Composition Explorer (ACE)
spacecraft, which is used to monitor the solar wind near the
Earth, never saw fluxes above 200 particles cm$^{-2}$ s$^{-1}$
ster$^{-1}$ MeV$^{-1}$ in the P3 channel ($112-187$~keV).
$Chandra$ observations are generally terminated when this rate
exceeds 50,000 to prevent damage to the ACIS CCDs. All particle
detectors associated with the on-board Electron Proton Helium
Instrument (EPHIN) also reported some of their lowest background
measurements during this period. Nonetheless some data were lost.
The typical ONC observation included about 45,000 frames (3.2-sec
exposures) from each of the five CCDs. Typically less than 10
frames (20 events) from each CCD were lost due to telemetry
errors. However the final observation, ObsID 3498, lost about 40
frames ($\sim$ 80 events) of the roughly 18,000 sent from each
CCD.

\section{Data Selection}\label{data_selection_section}

The COUP data reduction and cataloguing procedure is diagramed in
Figure \ref{diagram_fig}. We outline in this section the first
steps of data selection; see Appendix B of \citet{Townsley2003}
for more details on many of these steps. Procedures were performed
with {\rm{\it CIAO}} software package version 2.3, {\rm{\it
FTOOLS}} version 5.2, {\rm{\it XSPEC}} version 11.2, the Penn
State CTI corrector version 1.16, and the {\rm{\it acis\_extract}}
package version 2.33. The latter two tools were developed at Penn
State\footnote{Descriptions and codes for CTI correction and
{\rm{\it acis\_extract}} can be found at
\url{http://www.astro.psu.edu/users/townsley/cti/} and
\url{http://www.astro.psu.edu/xray/docs/TARA/ae\_users\_guide.html},
respectively.}.

Data reduction starts with Level 1 event files provided by the
$Chandra$ X-ray Center (CXC).  First, we filter out events on the
S2 chip. Second, likely cosmic ray and associated afterglow events
are identified, but not removed, by taking advantage of the Very
Faint telemetry mode.  Third, employing the PSU CTI Corrector
\citep{Townsley2002}, the data are partially corrected for CCD
charge transfer inefficiency (CTI) caused mainly by radiation
damage at the beginning of the $Chandra$ mission. Fourth, the data
are cleaned with filters on event ``grade'' (only ASCA grades
0,2,3,4,6 are accepted), ``status,'' and ``good-time interval''
filters supplied by the CXC. Energies $< 500$ eV and background
events with energies $> 10500$ eV are removed. Bad pixel columns
with energies $< 700$ eV, left by the standard processing, are
removed. Fifth, for each of the six COUP observations, individual
correction to the absolute astrometry was applied based on several
hundred matches between the preliminary $Chandra$ and VLT point
source catalogs (see \S \ref{stellar_counterparts_section}). Along
with reapplying the new aspect solutions to the data lists, the
slight broadening of the PSF from the CXC's software randomization
of positions was removed. Sixth, the tangent planes (x,y
coordinate systems) of the five COUP observations were
re-projected to match the tangent plane of the first observation
(ObsId 4395), and all six files were merged into a single data
event file.

For the purpose of source detection only, we eliminated flaring
pixel events and events identified earlier using Very Faint mode
that, when combined with random background events, can produce
faint spurious sources. This cleaning operation reduced the
background by $\sim 17\%$ but at the same time it removed $\sim
7\%$ of real photons from many bright sources.  In the later
source photon extraction stage (see Figure \ref{diagram_fig}), we
use an event list that retains these events so that source
brightnesses are not underestimated.

\begin{figure}
\centering
\includegraphics[width=1\textwidth]{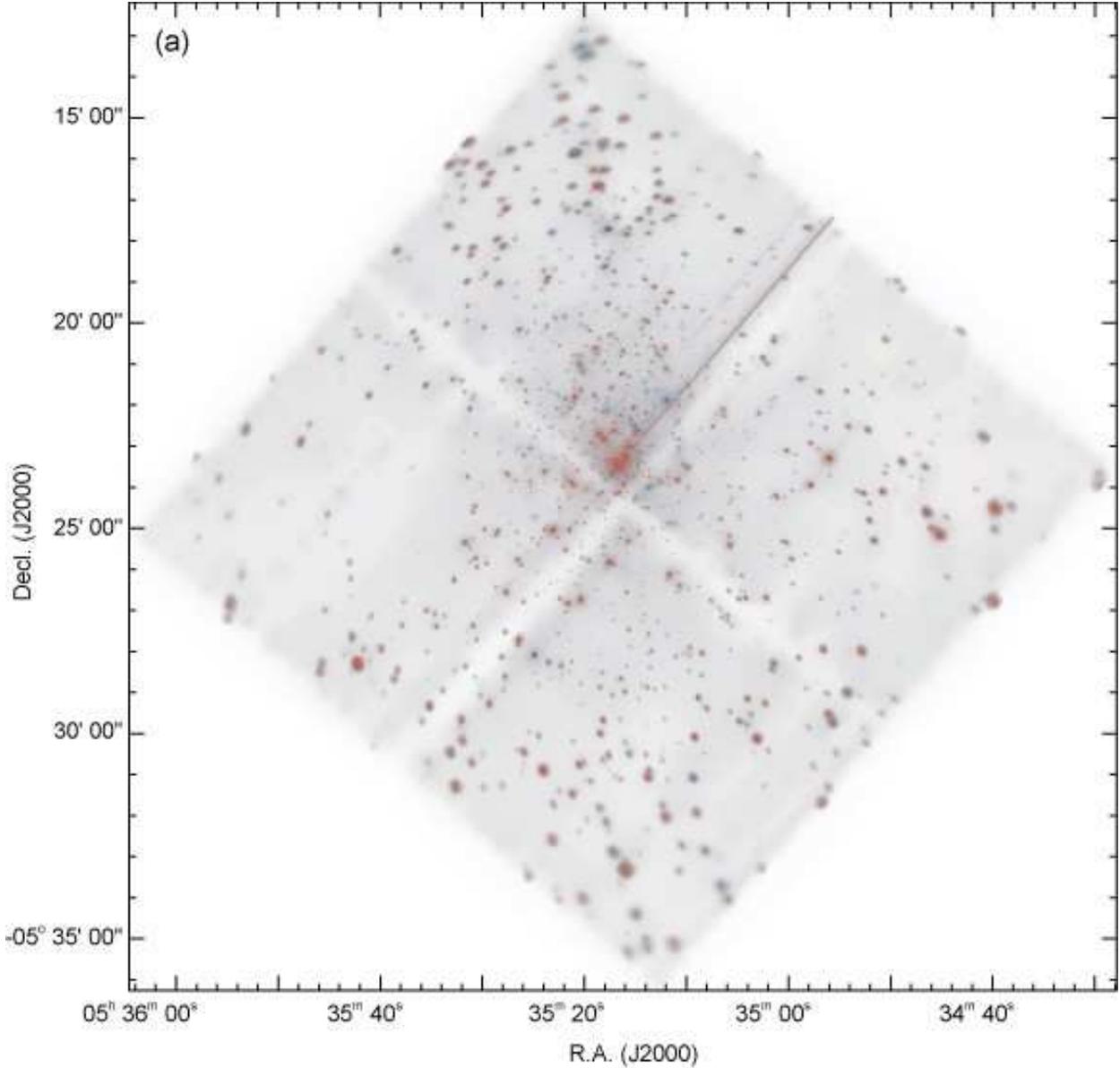}
\caption{The COUP ACIS-I image: (a) the full ACIS-I field shown
with 1\arcsec\/ pixels; and (b) the central region ($4\arcmin
\times 4\arcmin$) shown with 0.25\arcsec\/ pixels. The image is
shown after adaptive smoothing. The energy of each pixel is coded
in color so that soft ($0.5-1.7$ keV) (unabsorbed) sources appear
red while hard ($2.8-8.0$ keV), often absorbed, sources appear
blue. Green indicates moderately absorbed sources with typical
energies of $1.7-2.8$ keV. Brightness is scaled to the logarithm
of the photon number in the displayed pixel. The color model
depicts zero flux as white. Images are not corrected for the
exposure map and thus show gaps between the four CCD chips.
\label{ACIS_img_fig}}
\end{figure}

Figure \ref{ACIS_img_fig} shows the resulting images. Note the
bright CCD readout streaks emanating from the brightest sources.
In actuality, every source in the field produces similar readout
streaks, producing a spatially variable background which must be
considered when sources are detected. The chip gaps and bad pixel
columns in the form of white stripes are also noticeable, as the
images displayed here have not been corrected for the exposure
map.

\begin{figure}
\centering
\includegraphics[width=0.6\textwidth]{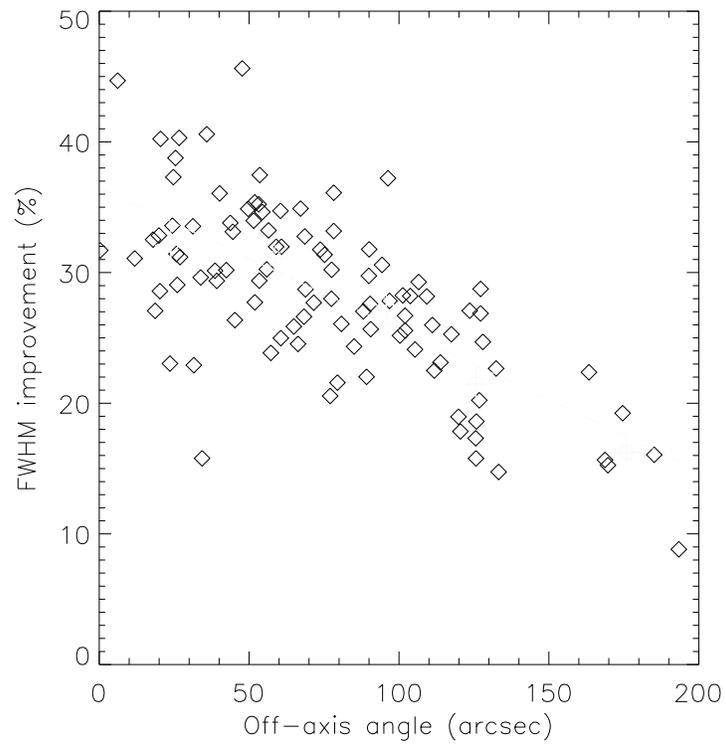}
\caption{Improvement in the full width half-maximum (FWHM) width
of the point spread function of 100 COUP sources due to
application of the subpixel repositioning (SER) procedure.
\label{SER_example_fig}}
\end{figure}

As a large fraction of PMS stars are expected to be in binary
systems, some with separations comparable to the $Chandra$ PSF, we
have applied a subpixel event repositioning (SER) technique to
improve the effective resolution of the ACIS detector. The SER
procedure used here considers the ``splitting'' of charge between
adjacent CCD pixels in an energy-dependent fashion \citep{Li04}.
We have applied SER to the merged exposure with cosmic ray
afterglow events removed. We tested the results of SER by
examining image quality for 100 COUP sources with $> 300$ counts
located at off-axis angles $\theta \lesssim 3\arcmin$. A 36\%
improvement in the PSF full width at half maximum (FWHM) is
achieved on-axis. As expected from the PSF broadening off-axis,
the SER-induced improvement decreases to 31\%, 21\%, and 16\% at
$\theta = 1\arcmin$, $2\arcmin$, and $3\arcmin$ off-axis,
respectively (see Figure \ref{SER_example_fig}).

\section{Source Detection \label{source_detection_section}}

As there is no single optimal procedure for identifying sources in
such a rich $Chandra$ field, trial source lists were prepared
independently by two groups within the COUP team. The Penn State
group used a wavelet transform detection algorithm implemented as
the {\rm{\it wavdetect}} program within the CIAO data analysis
system \citep{Freeman02}. The Palermo group used the Palermo
wavelet transform detection code, {\rm{\it PWDetect}}\footnote{
Description and code for {\rm{\it pwdetect}} are available at
\url{http://www.astropa.unipa.it/progetti\_ricerca/PWDetect}. For
the COUP analysis, the code was modified to accommodate the large
number of events, the conspicuous PSF wings due to bright sources,
and events occurring during CCD readouts.}. Source detection
procedures were carried out on both the original and SER-corrected
images, and on images constrained to include only soft and only
hard events.

\subsection{Penn State Procedure}\label{PSU_procedure_section}

A $Chandra$ imaging effective exposure map with units of (cm$^2$ s
counts photons$^{-1}$) was made from the product of an aspect
histogram {\rm{\it asphist}} and an instrument map. The aspect
histogram gives the amount of time the $Chandra$ optical axis
dwelled on each part of the sky, and is derived from the satellite
aspect solution. The instrument map provides the instantaneous
effective area across the field of view.  It includes the detector
quantum efficiency, non-uniformities across the face of a
detector, mirror vignetting, and bad pixels.

To reduce unnecessary computation on high-resolution images far
off-axis where the PSF is degraded, we constructed four $1024
\times 1024$ pixel scenes around the aimpoint with $0.25 \arcsec$,
$0.50 \arcsec$, $1.00 \arcsec$, and $1.44 \arcsec$ pixels. An
effective exposure map was created for each scene assuming a
monochromatic incident spectrum with photon energy of 1.6 keV, the
approximate energy at which the largest number of photons was
detected. To improve sensitivity to faint sources with unusually
soft or hard spectra, for each scene we produced energy
band-limited images from the merged event list (cleaned of flaring
events) in the total band ($0.5-8.0$ keV), soft band ($0.5-2.0$
keV), and hard band ($2.0-8.0$ keV).  These images were produced
from the original merged event list, for the event list with
background removal based on Very Faint telemetry mode, and for the
SER-corrected event list.  Source detection was thus performed on
36 $1024 \times 1024$ images: three event lists, four spatial
resolutions, and three spectral bands.

We ran {\rm{\it wavdetect}} with the probability threshold $P =
10^{-6}$ which theoretically permits a few false positive
detections in each image, and with $P=10^{-5}$ which permits
dozens of false positives. The wavelet scales were set between
1--2, 1--4, 1--8, and 1--16 pixels for the 0.25\arcsec,
0.50\arcsec, 1.00\arcsec, and 1.44\arcsec\ scenes respectively.
We then merged the catalogs, keeping only a single entry for each
source by removing duplicate entries obtained from detections in
lower-resolution images, if such entries existed. The sources were
then visually examined, resulting in the rejection of 184 {\rm{\it
wavdetect}} sources. Most of these were spurious triggers lying on
the readout trails or PSF wings of $\theta^1$C Ori or other very
bright sources. The remainder appear to be random associations of
events that did not resemble the local PSF. During this
examination of the image 109 closely spaced and weak but likely
sources were added to the PSU source list. The resulting combined
and cleaned {\rm{\it wavdetect}} source list has 1622 detections.

\subsection{Palermo Procedure}\label{Palermo_procedure_section}

{\rm {\it PWDetect}} is the $Chandra$ version of {\rm {\it
wdetect}}, a source detection code originally developed for\\
$ROSAT$/PSPC data, based on wavelet transforms
\citep{Dam97a,Dam97b}. It works directly on event files, instead
of binned images, and computes wavelet transforms on a range of
spatial scales, to be sensitive to point-like as well as
moderately extended sources. Compared to the $ROSAT$ version, the
code has been improved in order to deal with $Chandra$'s extremely
good spatial resolution, which implies a very sparse distribution
of background counts around a source.  This Poisson character of
the background was taken into account in establishing the
detection significance, using empirical formulae based on
simulations as described in \citet{Dam97a}. In the special case of
COUP data, {\rm {\it PWDetect}} was further refined to deal with
the very large number of events, and, more importantly, with the
presence of readout trails and conspicuous PSF wings, both
highlighted by the presence of bright sources and responsible for
scores of spurious detections if not accounted for. Readout trails
are now automatically identified and statistically removed prior
to detection, thus greatly reducing the number of spurious
detections. The wings of bright sources are treated by suitably
raising the detection threshold along the PSF wings. In this way,
we are able to discriminate real from false detections even very
close to the bright Trapezium stars.

For the COUP analysis, {\rm {\it PWDetect}} was run on the merged
event file restricted to events with energies between 300 and 7000
eV (the method was not applied separately to hard and soft bands)
and after application of the VF-mode background filtering. The
significance threshold was set at 5.0\,$\sigma$. This number is
not strictly the same as the signal/noise ratio, but is defined as
the probability of source existence, expressed in Gaussian sigmas.
Simulation of source-free background fields indicate that, at the
chosen 5.0\,$\sigma$ significance level, $\approx$10 spurious
sources should be present due to Poissonian fluctuations. This is
a good compromise: setting the threshold more conservatively at
5.5\,$\sigma$, corresponding to $\approx$1 spurious detection,
would result in detecting about 50 fewer sources, with a net loss
of $\approx$40 real sources.

Our initial list of {\rm{\it PWDetect}} sources for the original
event list contained 1497 COUP sources. From visual examination,
62 sources were rejected along readout trails, 5 were rejected in
the PSF wings of $\theta^1$C Ori, and 7 were flagged as tentative
due to their weakness. Fifteen additional sources were clearly
visible by eye but were missed by {\rm{\it PWDetect}} due to
proximity to other sources.  This particularly occurs when two
close sources far off-axis, where the PSF is elongated, mimic a
single radially symmetric source.  The same procedure applied to
the SER-corrected event file produced an initial list of 1493
sources. Seven of these were deemed legitimate new sources and
were added to the list. The final Palermo source list comprises
1452 detected sources.

\subsection{Final COUP Source List}\label{source_list_section}

From 64 sources detected by {\rm{\it wavdetect}} but undetected by
{\rm{\it PWDetect}}, 38 have significance greater than
$5.0\,\sigma$ (i.e., significance threshold for {\rm{\it
PWDetect}}). Of those 38, 19 are double sources, and another 19
are located at the regions of inhomogeneous exposure map (i.e.,
chip gaps, field edges, and bad pixel columns). The Palermo
procedure emerged with 10 sources missed by the Penn State
procedure, so the combined list has 1632 detections. Following
photon extraction and local background determination for each
source (\S \ref{photon_extraction_section}), 16 of these sources
were found to have $< 3$ net (i.e., background-subtracted) counts.
We subjectively decided that such weak sources are unreliable and
have placed them in a separate list of tentative sources.  The
resulting COUP source list of 1616 sources is presented in Table
\ref{COUP_src_table}.

The column entries to Table \ref{COUP_src_table} are as follows.
Column 1 gives the running source number which will be used in
this and other COUP papers. Sources are listed in order of right
ascension. Column 2 gives the source name in IAU format
Jhhmmss.s$-$ddmmss with the IAU designation CXOONCJ ($Chandra$
X-ray Observatory Orion Nebula Cluster J2000). Columns 3 and 4
give source position for epoch J2000.0 in degrees. For sources
with $\theta < 5\arcmin$, the positions are obtained from a simple
centroiding algorithm within the {\rm{\it acis\_extract}}
procedure. For sources located farther off-axis, the wavelet
positions were improved by correlating the event distribution with
PSF images using a matched-filter technique implemented by
{\rm{\it acis\_extract}}. Column 5 gives each source's positional
uncertainty in arcseconds. These are calculated within {\rm{\it
acis\_extract}} as 68\% (1\,$\sigma$) confidence intervals using
the Student's $T$ distribution\footnote{The Central Limit Theorem
\citep[see i.e., p.17 in][]{Lyons91} says that the mean values of
all possible samples of a given size N drawn from a particular
population tend to cluster in a bell-shaped curve (like Gaussian)
around the mean of the population from which the samples are
drawn, and the standard deviation of such a set of sample means
($\sigma_{\overline{x}}$) is related to the standard deviation of
the population ($\sigma$) as $\sigma_{\overline{x}} =
\sigma/\rm{\sqrt{N}}$. The positions of N events extracted from
the source region may be considered as a sample of size N out of
the set of all possible samples of that size drawn from the parent
population (i.e., a highly populated PSF, truncated by the
extraction region). In the case where the parent population
standard deviation is not known, a roughly 68\% confidence
interval for the population mean (i.e., statistical uncertainty on
source position) would be $1.0 \times \rm{S/\sqrt{N}}$, where S is
the observed sample standard deviation. This is true for large N.
For small samples the confidence interval becomes $t \times
\rm{(S/\sqrt{N})}$ where $t$ is the confidence level of Student's
$T$ distribution, whose exact shape depends on N \citep[see i.e.,
p.86 in][]{Lyons91}. For an infinite number of N the Student's $T$
distribution is the same as the Gaussian distribution.}. Column 6
gives the off-axis angle for each source in arcminutes.

Column 7 gives a flag indicating a variety of warnings and
difficulties. A ``u'' denotes uncertain objects; these are very
weak sources without stellar counterparts which satisfy the
detection criteria but are subjectively not completely convincing.
Seventy-four sources have ``u'' designations. A ``d'' refers to a
double source, defined as two sources with overlapping 90\% PSF
contours.  A ``p'' defines a possible piled-up source, when the
surface brightness of the source exceeds a level of 0.003 counts
s$^{-1}$ pixel$^{-1}$. A ``t'' indicates that a source region
crosses a bright source readout trail. A ``w'' indicates that the
source is located in the wings of any bright source with number of
counts $>20000$. An ``x'' indicates a region of inhomogeneous or
low exposure map, where the source is located near or on chip
gaps, bad pixel columns, or field edges.  Sources designated ``x''
will have disturbed lightcurves.

Column 8 defines the type of the source detection. The first digit
indicates the image on which {\rm{\it wavdetect}} found the
source: 1 = total band image, 2 = soft band image only, 3 = hard
band image only, and 4 = total band image with background filtered
in Very Faint mode only. The great majority of sources (1522 of
1616) were found in the total band image; still 45 sources were
found only in the soft band, and 36 only in the hard band. The
second digit indicates the type of event list on which a source
was detected using {\rm{\it PWDetect}}: 1 = merged event list
filtered in Very Faint mode, and 2 = merged SER corrected event
list.

Column 9 indicates parameters of the detection procedure. The
first digit refers to {\rm{\it wavdetect}} where 1 = probability
threshold $P = 1\times 10^{-6}$ and 2 = probability threshold $P =
1 \times 10^{-5}$. A ``1'' in the second digit indicates that the
source was detected with the {\rm{\it PWDetect}} algorithm only.
A non-zero entry in the third digit indicates that the source was
found only by visual inspection. Here, ``1'' sources were found by
both the Penn State and Palermo groups, ``2'' sources were found
only by the Penn State group, and ``3'' sources were found only by
the Palermo group.  A total of 123 of the 1616 sources were found
only by visual inspection.

Column 10 gives the source significance calculated by {\rm{\it
acis\_extract}}, which indicates how many times the source's net
(background-subtracted) counts exceed the uncertainty on that
quantity ${\rm SigAE = NetCts/\sigma(NetCts)}$ (\S
\ref{photon_extraction_section}). Column 11 defines the
probability of source existence, expressed in Gaussian sigmas
calculated by {\rm{\it PWDetect}}.

Columns 12 and 13 list the source counterparts from previous
$Chandra$ studies with the ACIS \citep{Feigelson02a} and HRC
\citep{Flaccomio03a} detectors.  Counterparts to earlier
$Einstein$, $ROSAT$, and $ASCA$ sources are given by
\citet{Feigelson02a}.

Table \ref{tentative_table} lists the 16 tentative sources which
passed the Penn State and Palermo source detection criteria but
did not have $>3$ net counts after photon extraction.  Here we
give only positions. These possible sources are not discussed
further in this paper. The photometry results for those very weak
sources strongly depend on the way local backgrounds are chosen.
By introducing manually selected local background (\S
\ref{photon_extraction_section}), we may improve their photometry
and reconsider our preliminary decision on whether these are real
or tentative sources. This work will be done in the future COUP
membership paper, Getman et. al. (in preparation).

\section{Photon Extraction}\label{photon_extraction_section}

The {\rm{\it acis\_extract}} package is used to extract source
photons, refine the accuracy of source positions, estimate local
background, construct source and background spectra, compute
redistribution matrix files (RMFs) and auxiliary response files
(ARFs), construct lightcurves and time-energy diagrams, perform a
Kolmogorov-Smirnov variability test, compute photometric
properties, and perform automated spectral grouping and fitting.
The reader is referred to the {\rm{\it acis\_extract}} Users
Guide\footnote{Users Guide for {\rm{\it acis\_extract}} can be
found online at
\url{http://www.astro.psu.edu/xray/docs/TARA/ae\_users\_guide/.}}
for further details on the implementation of each step. The
quantities derived in the photon extraction stage are listed in
Table \ref{COUP_pho_table}, and we refer to the columns of this
table in this section.

Selection of an optimal extraction region for each source is not a
simple task; one seeks a balance between a larger region
maximizing the source's signal and (for weak sources) a smaller
region minimizing the background's signal.  The size and shape of
the extraction region vary greatly over the field of view and
cannot be parametrized accurately by an analytical expression.
{\rm{\it Acis\_extract}} treats these difficulties by extracting
the events around the source centroid (initially, the {\rm{\it
wavdetect}} position) inside a polygonal contour of the local PSF,
obtained with the CIAO tool {\rm {\it mkpsf}}.  The user need only
specify the fraction of the enclosed energy desired for the
extraction. For most sources, we chose to extract events from the
polygonal contours of $\sim 87\%$ encircled energy using PSFs at
the fiducial energy of 1.497 keV. For $\sim 250$ weak and crowded
COUP sources, we chose smaller PSF contours ranging from 20\% to
80\% encircled energy.

While the ACIS-I instrumental background level is usually
spatially invariant, it can change substantially across the COUP
field due to the PSF wings and readout trails of the strong
sources. We therefore adopt the procedure used in {\rm{\it
acis\_extract}} for automatically measuring and subtracting a
local background individually for each source. The background
extraction region starts with a circular annulus where the inner
radius circumscribes the $\sim 1.1 \times 97\%$ PSF polygon and
the outer radius is set such that the background region
accumulated $> 100$ events. These background events are obtained
from a special background image that has excluded all events
within the $\sim 1.1 \times 97\%$ PSF circles of all 1616 sources,
so background events for one source are not corrupted by source
events from neighboring sources.

\begin{figure}
\centering
  \includegraphics[angle=0.,width=0.9\textwidth]{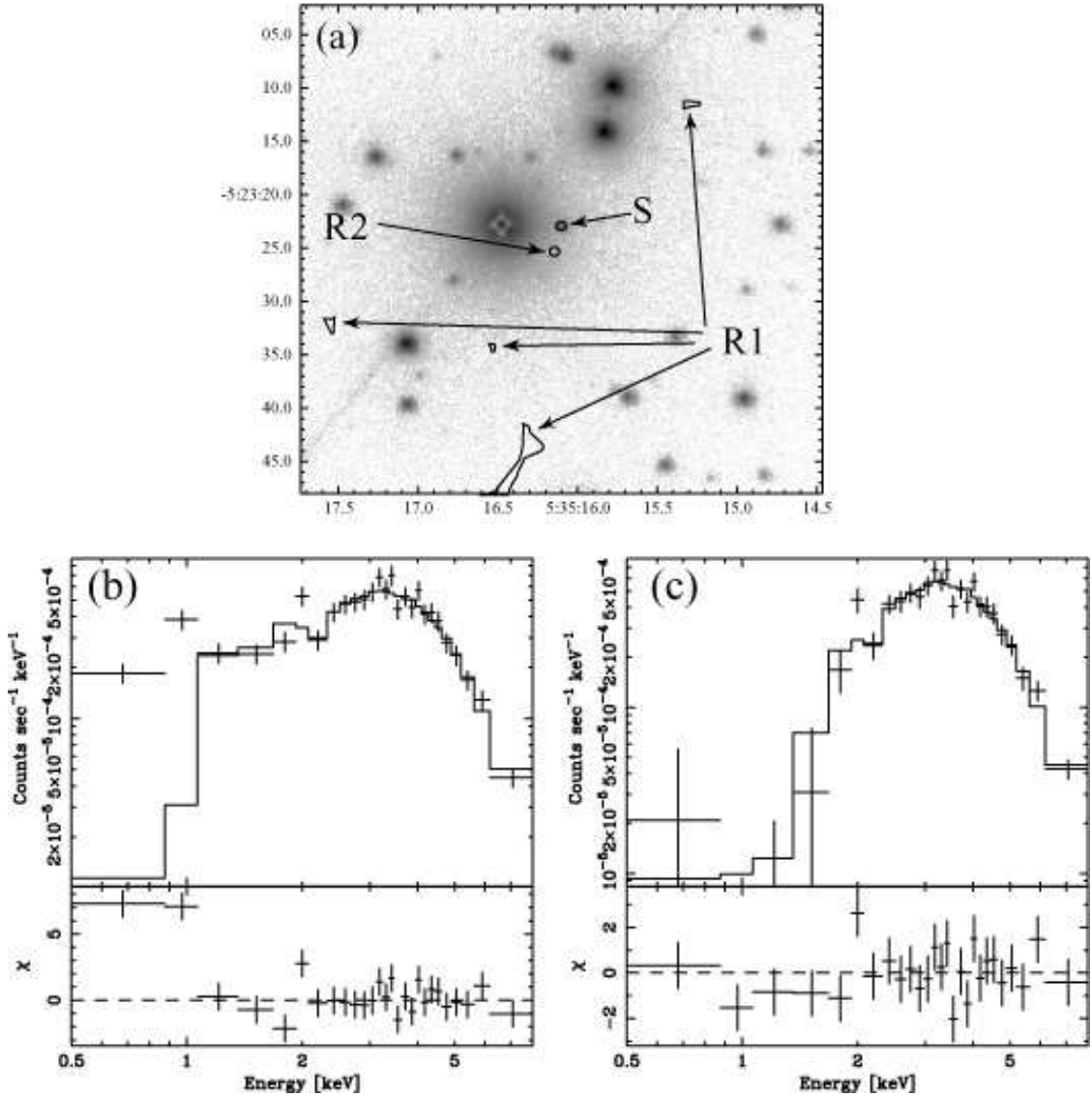} \hspace{0.05in}
\caption{(a) Closeup view of the neighborhood of the sample source
\#780, located in a region of highly non-uniform background. R$_1$
marks the automatically extracted background region, R$_2$ is the
manually improved local background, and S indicates the source
extraction region. (Bottom) Spectra obtained through the
subtraction of the automatically extracted background (b), and
improved manually chosen background (c). \label{badimg_srcs_fig}}
\end{figure}

{\rm{\it Acis\_extract}} can also accept a user-generated
background to accommodate unusual sources.  The local background
region was manually adjusted for 57 weak sources located in
crowded fields, near or on readout trails, or on PSF wings of
bright sources. Improvement of the local background is shown in
Figure \ref{badimg_srcs_fig} for the sample COUP source \#780. The
spectrum to the left is extracted with the automatically
subtracted background $R_1$, while the spectrum to the right is
more realistic, and obtained by subtracting the manually chosen
background $R_2$. Even with such care, some faint sources may
still have spectra that are corrupted by incorrect background
subtraction.

Results from the photon extraction procedure appear in Table
\ref{COUP_pho_table}. The source counts tabulated in Column 3 give
the number of counts in the total (0.5 - 8~keV) energy band
extracted for each source (including background). Column 4 gives
the number of background counts in the total band scaled to the
extracted area corrected for small differences that may be present
in the exposure maps of the source and background regions. The net
or background-corrected counts appear in Column 5; this is the
difference between Columns 3 (extracted source counts) and 4
(scaled local background counts). Columns 6 and 7 give the area of
the extraction polygon in $0.5\arcsec \times 0.5\arcsec\/$ pixels
and the fraction of the PSF at the fiducial energy of 1.497 keV
enclosed within the extracted area.

Spectral analysis requires calibration files (ARFs and RMFs)
specifying the effective area and spectral resolution of the
instrument at each source location as a function of energy.  As
the Penn State CTI corrector has been applied to the COUP event
data (\S \ref{data_selection_section}), there is no need to
calculate RMFs for each source individually. Given the source
location on each CCD chip, we select the appropriate RMF from a
standard suite provided with the CTI corrector.

Source-specific ARF files are then calculated using the energy
grid specified in these RMFs (685 PI channels, or 410 energy
bands). In cases where a source spans multiple CCDs due to
satellite dithering, the ARFs were calculated for both CCDs and
summed. A further correction to the ARFs is applied to account for
the energy dependence of the source PSF that falls outside the
extraction region. PSFs were constructed at five fiducial energies
between 0.27 and 8.6 keV (0.277, 1.497, 4.510, 6.400, and 8.600
keV), the polygonal source extraction region was applied to each
PSF, the PSF fraction was computed at each available energy, and
the ARF was reduced by this PSF fraction curve. The ARF was also
corrected for the hydrocarbon build-up on the ACIS filters using
the correction curve supplied by the tool {\rm{\it
acisabs}}\footnote{A description of the {\rm{\it acisabs}}
procedure can be found online at
\url{http://www.astro.psu.edu/users/chartas/xcontdir/xcont.html}.}.

The ``effective'' exposure time at the source location is given in
Column 8. This quantity is derived by normalizing the ratio of an
average exposure map (``Ct\_Fl'') to the one obtained for the
region with the maximum value of the exposure map. Specifically,
${\rm Exp} = {\rm (Ct\_Fl / Ct\_Fl_{max})} \times 838$~ks, where
838 ks is the COUP ``observed'' exposure time (sum of all
exposures in Table \ref{observation_journal_table}). This quantity
summarizes the variation in the ``depth'' to which the sources
were observed.

Column 9 gives the value of the exposure map averaged over the
source region (Ct\_Fl) in units of ($10^{9}$ counts s cm$^{2}$
photon$^{-1}$) and represents a conversion factor between a photon
incident flux and detected counts.

The final eight columns of Table \ref{COUP_pho_table} provide a
variety of non-parametric measures of the source flux and
spectrum. Column 10 gives a rough estimate of the incident flux at
the telescope aperture in units of (photons cm$^{-2}$ s$^{-1}$):
${\rm IncFl = NetCts / <ARF> / Exp}$. Note that this flux estimate
becomes inaccurate when the spectrum is not flat over the total
$0.5-8$ keV band covered by the ARF. Summing the fluxes computed
over narrow energy bands will produce a more accurate result, and
parametric spectral fitting (\S \ref{spectral_analysis_section})
should produce the most accurate flux estimates. Column 11 gives
the background-corrected median photon energy MedE in the
$0.5-8.0$ keV range. IncFl and MedE can be used together for
estimating the luminosities of weak sources (say, ${\rm NetCts}
\lesssim 20$) for which nonlinear spectral fitting packages are
ineffective (\S \ref{luminosities_section}).

Columns $12-17$ describe three hardness ratios and their
$1\,\sigma$ upper and lower statistical errors. The hardness
ratios are defined by ${\rm HR = (Cts_h - Cts_s) / (Cts_h +
Cts_s)}$ where subscripts "h" and "s" refer to a harder and softer
band, respectively. For COUP, we define HR1 to reproduce the
commonly used hardness ratio between the ${\rm s = 0.5-2.0}$ keV
and ${\rm h = 2.0-8.0}$ keV bands, HR2 to highlight the softer
part of the spectrum between ${\rm s = 0.5-1.7}$ keV and ${\rm h =
1.7-2.8}$ keV, and HR3 to measure the harder part of the spectrum
between ${\rm s = 1.7-2.8}$ keV and ${\rm h = 2.8 - 8.0}$ keV. For
each energy band, {\rm{\it acis\_extract}} computes the net counts
corrected for the local background in the appropriate band.
Confidence intervals encompassing 68\% of the expected error,
equivalent to $\pm 1\,\sigma$ in a Gaussian distribution, are
estimated first by calculating upper and lower errors on total and
background counts using \citet{Gehrels86} equations (7) and (12),
and then propagating those errors to net counts using
\citet{Lyons91} equation (1.31). We then propagate errors from net
counts to upper and lower uncertainties on hardness ratios using
the method in \citet{Lyons91}. When net counts are negative in an
energy band we clip them at zero (to ensure that hardness ratios
are bounded by [-1,1]) and choose to set their lower errors to
zero. In cases where the 68\% confidence interval of soft band net
counts contains zero (very hard sources) we consider the upper
uncertainties on hardness ratios to be unreliable and do not
report them, similarly we do not report the lower uncertainties on
hardness ratios of very soft sources. When the 68\% confidence
interval of both bands contains zero counts we do not report
hardness ratios, because their errors would be large.

\section{Strong Sources with Photon Pile-up}\label{pile-up_section}

A few dozen COUP X-ray sources are simultaneously sufficiently
strong and close to the aimpoint with narrow PSFs to suffer photon
pile-up. This occurs when two or more photons are incident on a
single CCD pixel during a single 3.2 s CCD frame. Some of these
pile-up events mimic cosmic rays and are rejected by the on-board
computer, while others are telemetered as valid events but with
spuriously high energies.  While statistical reconstruction
treatments of ACIS pile-up have been developed \citep{Davis01,
Kang03}, the approach adopted here is to discard events at the
core of the point spread function, which are both spatially and
spectrally distorted, and to use only ordinary, pile-up-free
events in the outskirts of the PSF \citep{Broos98}.

\begin{figure}
\centering
  \includegraphics[angle=0.,scale=0.8]{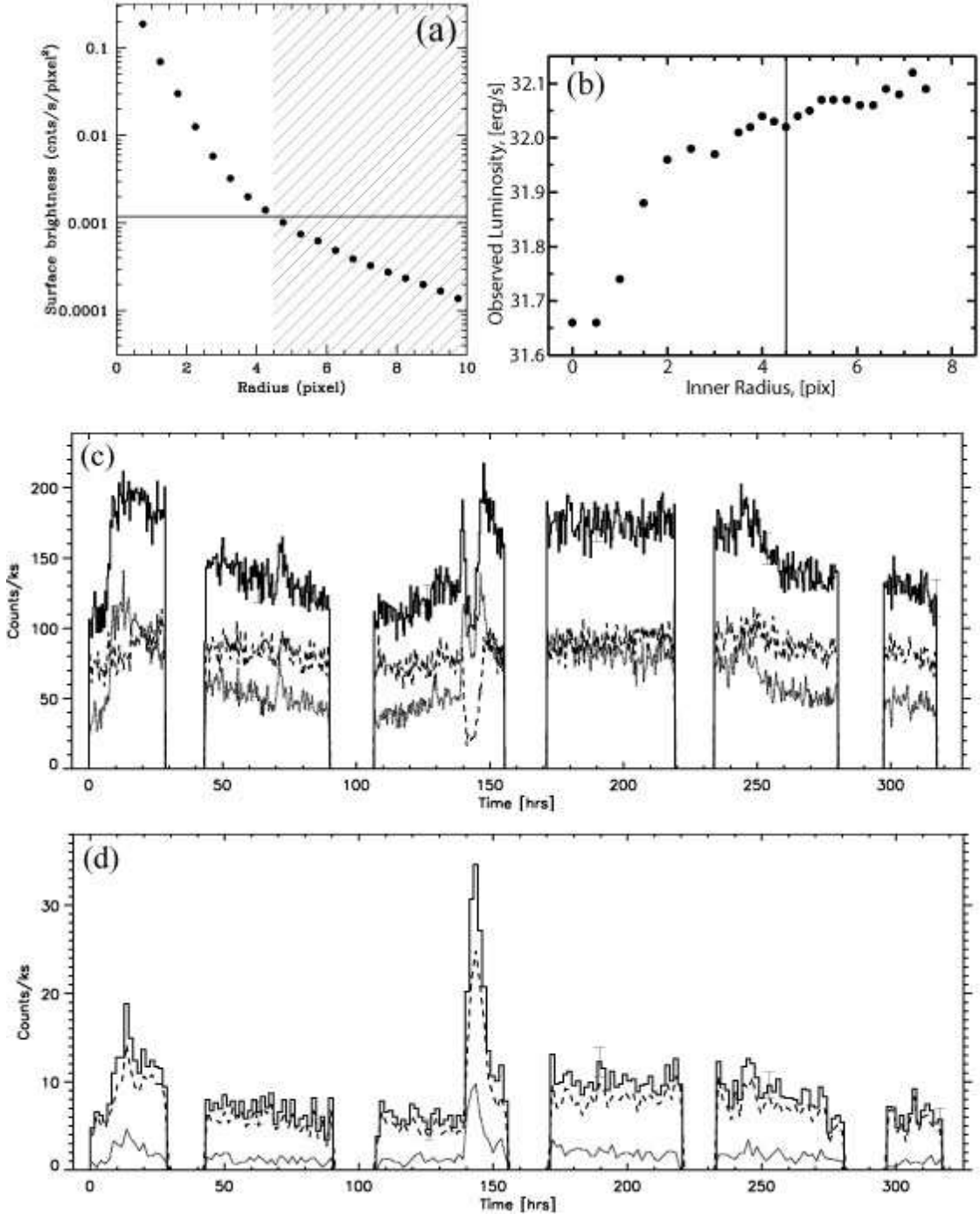} \hspace{0.05in}
\caption{Study of the heavily piled-up COUP source \#932: (a) the
radial surface brightness profile of the piled-up events; (b) the
  inferred source luminosity as a function $r_{in}$ in our annular
  extraction technique for a series of excluded core sizes (the solid
line marks the $\sim 2\%$ pile-up fraction); (c) lightcurve
extracted from the usual extraction region including pile-up; and
  (d) lightcurve from the optimal annular extraction region. In the
  lightcurves, the solid line indicates the total energy band
  ($0.5-8$ keV), dashed line indicates the soft band ($0.5-2.0$
  keV), and dotted line indicates the hard band ($2.0-8.0$ keV).
  Note that panel (c) shows a spurious dip while panel (d) shows the
correct flare.\label{pile-up_fig}}
\end{figure}

Our procedure is to extract events from circular annuli between
radii $r_{in}$ and $r_{out}$ (the latter was usually fixed around
the 99\% enclosed energy contour) and to construct specialized
ARFs associated with those annuli at the particular source
location.  The ARF correction is made using the {\it xpsf} tool
developed by G.\ Chartas\footnote{Description of the procedure
based on the algorithm of {\it xpsf.pro} can be found online at
\url{ http://www.astro.psu.edu/users/tsujimot/arfcorr.html}.}
which uses MARX simulated sources to establish the
energy-dependent fraction of photons in the extracted annulus. The
challenge is to find the smallest $r_{in}$ which gives the
greatest number of counts available for later analysis without
reducing the inferred source flux because of photon pile-up. The
best $r_{in}$ was established for each source by repeated trials
\footnote{More information can be found at \url{
http://www.astro.psu.edu/xray/docs/TARA/ae\_users\_guide/pileup.txt}.}.
We found that the best $r_{in}$ typically occurs at a surface
brightness of $\sim 1.2 \times 10^{-3}$ counts s$^{-1}$
pixel$^{-1}$ which roughly corresponds to a pile-up fraction of
$\sim 2\%$, or when the flux profile reaches a plateau. This is
seen in the top two panels of Figure \ref{pile-up_fig} for the
heavily piled-up COUP source \#932.  The left panel (a) presents
the brightness profile with the dashed line marking the level of
$\sim 2\%$ pile-up fraction and the right panel (b) presents the
luminosity (flux ${\rm \times 4\pi D^{2}}$) profile obtained from
the spectral fits of consequently diminishing annuli.

Sixty-five strong and centrally concentrated COUP sources were
examined in this fashion; these sources are flagged ``p'' in Table
\ref{COUP_src_table}.  Only 24 sources were found to require the
annular extraction analysis with results given in Table
\ref{pile-up_table}. Those 24 piled-up sources are also flagged
with ``a'' in Table \ref{COUP_spe_table} (see \S
\ref{spectral_analysis_section} below). All further analysis for
them (photometry, spectroscopy, variability) was performed with
the annular extracted events.  It is particularly critical that
variability not be studied using the central piled-up core of
these sources: flares often saturate and can even appear as
plateaus or sudden dips in the count rates. An example of this
spurious result is shown in panels (c) and (d) of Figure
\ref{pile-up_fig}, where panel (c) presents the lightcurve
extracted from the usual extraction region including pile-up,
while panel (d) shows the lightcurve from the optimal annular
extraction region.

\section{Spectral Analysis} \label{spectral_analysis_section}

A goal of the present study is to produce an {\it acceptable}
spectral model for the 1616 sources to give reliable time-averaged
broadband luminosities and preliminary characterization of
intrinsic spectral properties and interstellar absorption. We
recognize that the semi-automated approach here using the {\rm{\it
acis\_extract}} package will often not produce the {\it best}
spectral model. Other COUP studies will analyze selected sources
in much more detail, studying interstellar absorptions, plasma
elemental abundances, time-variations of temperatures, and so
forth.  We describe the {\rm{\it acis\_extract}} spectral analysis
procedure here.

For each of the 1616 COUP sources, the CIAO tool {\rm{\it
dmextract}} is first called to create a HEASARC/OGIP-compatible
Type I source spectrum over the energy range $0-10$ keV, or {\rm
PI} channel range $0-685$ compatible with the Penn State
CTI-corrected RMFs. The {\rm BACKSCAL} keywords in the source and
background spectra are the integral of the exposure map over the
source and background regions. An algorithm for grouping spectra
was implemented using custom code inside {\rm{\it acis\_extract}},
rather than via the grouping tools in FTOOLS ({\rm{\it grppha}})
or CIAO (${\rm{\it dmgroup}}$), in order to place the first and
last grouping bin boundaries at specified energies so that the
spectra examined always have the range $0.5-8.0$ keV. These
boundaries are controlled by specifying ${\rm PI} = 1:34$ (energy
$<0.5$ keV) for the first group and ${\rm PI} = 549:685$ (energy
$\geq 8.0$ keV) for the last group. For the intermediate channels
${\rm PI} = 35:548$ of scientific interest, channels were grouped
according to the following scheme: each group contains at least 5
events for sources with ${\rm NetCts < 100}$, at least 7 events
for ${\rm 100 \leq NetCts <200}$, at least 10 events for ${\rm 200
\leq NetCts <500}$, at least 20 events for ${\rm 500 \leq NetCts
<1000}$, and at least 60 events for ${\rm NetCts \geq 1000}$.
Simulations show that biases to the inferred temperatures when $kT
\geq 10$ keV and to the column densities when $\log N_H \leq 20.5$
cm$^{-2}$ may be present when  groups have ${\rm NetCts \leq 10}$
events \citep[][~\S 2.8]{Feigelson02a}.

Automated spectral fitting is performed by {\rm{\it
acis\_extract}} by spawning the {\rm{\it XSPEC}} spectral fitting
program \citep{Arnaud96}. In most cases, spectra fits were found
with one- and two-temperature optically thin thermal plasma MEKAL
models \citep{Mewe91}, assuming a uniform density plasma with 0.3
times solar elemental abundances \citep{Imanishi01,Feigelson02a}.
While these fits are usually statistically successful, the
assumptions are clearly astrophysically inadequate. Detailed
studies of magnetically active stars show that many plasma
components with a wide range of temperatures may be present, and
that chemical fractionation associated with magnetic reconnection
flaring can produce strongly non-solar elemental abundance
distributions \citep[e.g.][]{Kastner02, Audard03}.  X-ray
absorption was modelled using the atomic cross sections of
\citet{Morrison83} with traditional solar abundances to infer a
total (mainly hydrogen and helium) interstellar column density,
$\log N_H$, in units of (atoms cm$^{-2}$)\footnote{Three strong
COUP X-ray sources with complex spectra required two absorption
components to achieve an acceptable spectral fit (in {\rm{\it
XSPEC}}, this model is specified $wabs \times mekal + wabs \times
mekal$). These column densities were, for the three sources
respectively: $\log N_{H1} < 20$ and $\log N_{H2} = 22.3$ for COUP
source \#267; $\log N_{H1} = 21.6$ and $\log N_{H2} = 22.8$ for
\#578; and $\log N_{H1} \sim 20.9$ and $\log N_{H2} \sim 22.8$ for
\#948. These values are not reported in Table
\ref{COUP_spe_table}.  It is not clear whether the two components
are astrophysically real (e.g., an unresolved double source with
different absorptions) or result from poor spectral model
specification. \label{two_abs_footnote}}. Here again, the
astrophysical situation may be more complex with untraditional
elemental abundances or gas-to-dust ratios \citep{Vuong03}.

Acceptable spectral fits were obtained by minimizing $\chi^2$
between the grouped ACIS spectra and the parametric plasma models.
We starting the fitting process in two ways: fixed standard values
of the model parameters ($\log N_H = 21.3$ cm$^{-2}$, $kT_1 = 1.0$
and $kT_2 = 3.0$ keV), and a grid of initial parameters. The final
result was chosen by visual inspection using {\rm{\it
acis\_extract's}} accessory tool {\rm{\it spectra\_viewer}}, based
on three criteria. First, for sources with ${\rm NetCts \lesssim
200}$, one-temperature plasma models were used if they fit.
Second, for hard sources with median energy ${\rm MedE \gtrsim 2}$
keV, a one-temperature model fit was again chosen to avoid the
highly nonlinear interaction of an unseen soft component and
uncertain interstellar absorption. Third, for the brighter and
softer sources  (${\rm MedE \lesssim 2}$ keV, ${\rm NetCts \gtrsim
200}$), two-temperature plasma fits were typically needed to
obtain good fits. Uncertainties to each spectral parameter are
estimated by a $\Delta \chi^2_{min} +$ constant criterion using
the {\it error} command in {\rm{\it XSPEC}}.

Emission measures were calculated using the best-fit
normalizations from XSPEC as\\ ${\rm EM = 10^{14} \times norm
\times 4\pi D{^2} }$. In cases of two-temperature fits, both
emission measures for soft and hard components are provided.

We emphasize again that the resulting spectral parameters and
their uncertainties are often not reliably determined, and that
alternative models may be similarly successful.  The greatest
uncertainty in plasma temperatures and emission measures occurs
for sources with high absorption, $\log N_H \simeq 22-23$
cm$^{-2}$. Here we have no direct knowledge of the soft emission
which, in less absorbed sources, typically dominates the hard
emission. Thus, our approach here may severely underestimate
broadband $0.5-8$ keV luminosities of heavily absorbed sources,
although luminosities in the hard $2-8$ keV band are more
reliable.

\begin{figure}
\centering
  \includegraphics[angle=0.,scale=0.6]{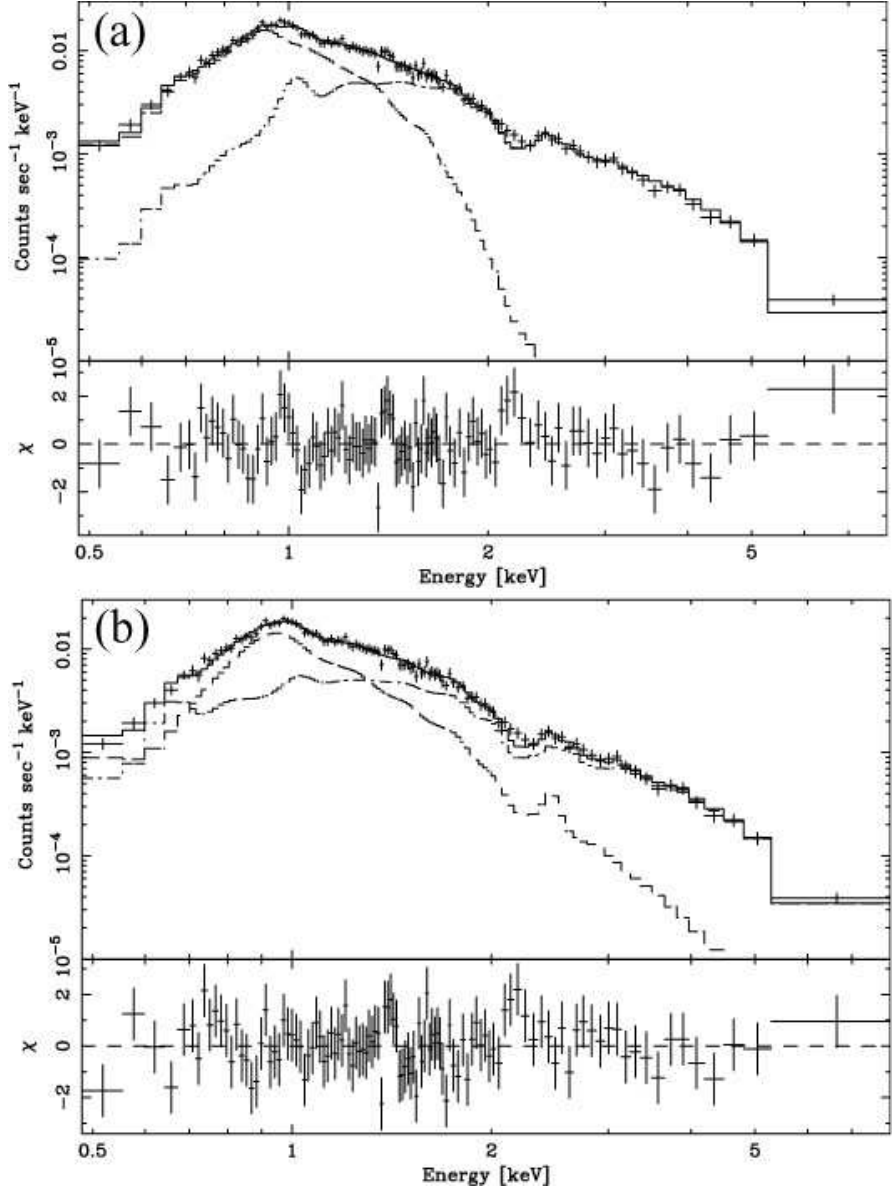} \hspace{0.0in}
\caption{Spectrum of COUP source \#567 exemplifying a common
ambiguity in spectral fitting.  Panel (a) presents a fit with
$kT_{1} \thickapprox 0.2$~keV, $kT_{2} \thickapprox 1.7$~keV,
$\log N_{H} \thickapprox 21.8$~cm$^{-2}$, $\log L_{t,c}
\thickapprox 31.6$~erg/s with a reduced $\chi^{2} \thickapprox
1.3$. Panel (b) presents another fit (which appears in Table
\ref{COUP_spe_table}) with $kT_{1} \thickapprox 0.8$~keV, $kT_{2}
\thickapprox 3.1$~keV, $\log N_{H} \thickapprox 21.3$~cm$^{-2}$,
$\log L_{t,c} \thickapprox 30.6$~erg/s, and a reduced $\chi^{2}
\thickapprox 1.4$. Note the 10-fold difference in inferred
unabsorbed luminosity in the total $0.5-8$ keV band.
\label{spectral_fit_problem_fig}}
\end{figure}

A second common source of uncertainty occurs for strong COUP
sources where two qualitatively different spectral models give
similarly good values of $\chi^{2}$.  Figure
\ref{spectral_fit_problem_fig} shows a typical example; the left
panel shows the solution with the lower energy $kT_{1}$ of
$0.1-0.3$ keV and higher energy $kT_{2}$ of $1-2$ keV, while the
right panel presents another solution with $kT_{1} \sim 0.6-0.9$
keV, $kT_{2} \sim 2-3$ keV. Such ambiguities in spectral
parameters occur in $\sim 150$ bright COUP sources.  We have
chosen here to present the second class of solution that avoids
the inference of a very luminous, heavily absorbed, ultra-soft
($kT_1 < 0.5$ keV) component. The errors in {\rm{\it XSPEC}}
parameters do not take into account the possibility that we have
chosen the incorrect acceptable model.

The results of our spectral analysis are tabulated in Table
\ref{COUP_spe_table}. One-temperature spectral fit results are
recorded for 980 COUP sources and two-temperature results for 563
sources. Spectral models for 73 sources with very poor statistics
(${\rm NetCts < 20}$) are left blank.  Columns 3 and 4 report the
logarithm of the hydrogen column density and its $1\,\sigma$ error
in cm$^{-2}$, obtained from the spectral fit. Fitted values with
$\log N_H < 20.0$ cm$^{-2}$ are truncated at 20.0 because ACIS-I
spectra are insensitive to differences in very low column
densities. Columns 5 and 6 give the plasma energy and its
$1\sigma$ error in keV; in cases of two-temperature fits, this is
the energy of the soft component. Columns 7 and 8 give the plasma
energy of the hard component and its $1\,\sigma$ error in
two-temperature model fits. Fitted values with $kT
> 15$ keV are truncated at 15 keV because the data can not discriminate
between very high temperature values. Columns 9-12 report the
emission measures and their $1\,\sigma$ errors in the units of
cm$^{-3}$ for each temperature component. Columns 13 and 14 give
the reduced $\chi^{2}_\nu$ for the overall spectral fit and
degrees of freedom, respectively.

Details regarding the spectral fitting process for individual
sources are provided by the flags in columns $15-16$. Column 15
marks sources where the model was formally inadequate, based on
the null hypothesis probability $P_{\chi}$ of $\chi^{2}$ for the
relevant degrees of freedom: ``m'' indicates a marginal fit with
$0.005 \leq P_{\chi} < 0.05$, and ``p'' a poor fit with $P_{\chi}
< 0.005$. Column 16 presents a three-part flag giving details of
the spectral modelling for each source. The first part gives the
minimum number of events in each spectral group. The second part
indicates whether two absorption components are needed (see
footnote \ref{two_abs_footnote}).  The third part gives the number
of plasma components used in the accepted fit, 1 or 2. A ``g''
indicates that the spectral result has been chosen from a grid of
initial parameter values.

Column 17 is a conjunction of eight flags giving important
information on source-specific spectral features and problems,
some obtained from visual inspection of the spectral fit and the
location of the source in the ACIS image. The first flag ``l''
indicates the presence of narrow spectral features in the data
that are not present in the model, probably due to elemental
abundances inconsistent with our assumption of $0.3~\times~$solar
values.  A total of 186 COUP sources are flagged with spectral
features.  Flags 2 and 3 indicate that soft (``s'') or hard
(``h'') excesses were present in the data that were not in the
model. In most cases, this can be attributed to poor subtraction
of nonuniform local background around weak sources. The fourth
flag ``c'' signals that the spectrum may be corrupted by a close
star component. The fifth flag ``p'' indicates that the spectral
fit is unreliable due to poor statistics and/or a poor fit based
on visual examination rather than $\chi^2$ values. The sixth flag
``m'' marks cases where we defined the background extraction
region with a manual (rather than automatic) procedure (\S
\ref{photon_extraction_section}).  The seventh flag ``a'' marks
the 24 heavily piled-up sources requiring annular extraction, as
discussed in \S \ref{pile-up_section}.  The eighth flag ``w''
notes 26 additional sources that appeared weakly piled-up for
which the usual whole polygonal extraction region was used.

\section{Variability Analysis} \label{variability_analysis_section}

The temporal behaviors of COUP sources are often very complex with
high-amplitude, rapidly changing flares superposed on quiescent or
slowly variable emission. Relegating detailed study of these
behaviors to later studies, we provide here two broadly applicable
measures of variability and report results in Table
\ref{COUP_var_table}. All variability measures are made on the
extracted SrcCts events without background subtraction. While the
background is usually negligible and is not itself highly
variable, it may play a role in the variability characteristics of
off-axis weaker sources. In 33 cases when sources were located on
an ACIS CCD chip gap (flagged in column 3 of Table
\ref{COUP_var_table}), tests have been performed on data extracted
from the primary CCD only. These sources may exhibit spurious
short-term variability.  We are confident that the intrinsic
variability of piled-up sources is accurately reflected by our
annular extraction procedure (\S \ref{pile-up_section}) though
with reduced signal.

\begin{figure}
\centering
  \includegraphics[angle=0.,scale=0.8]{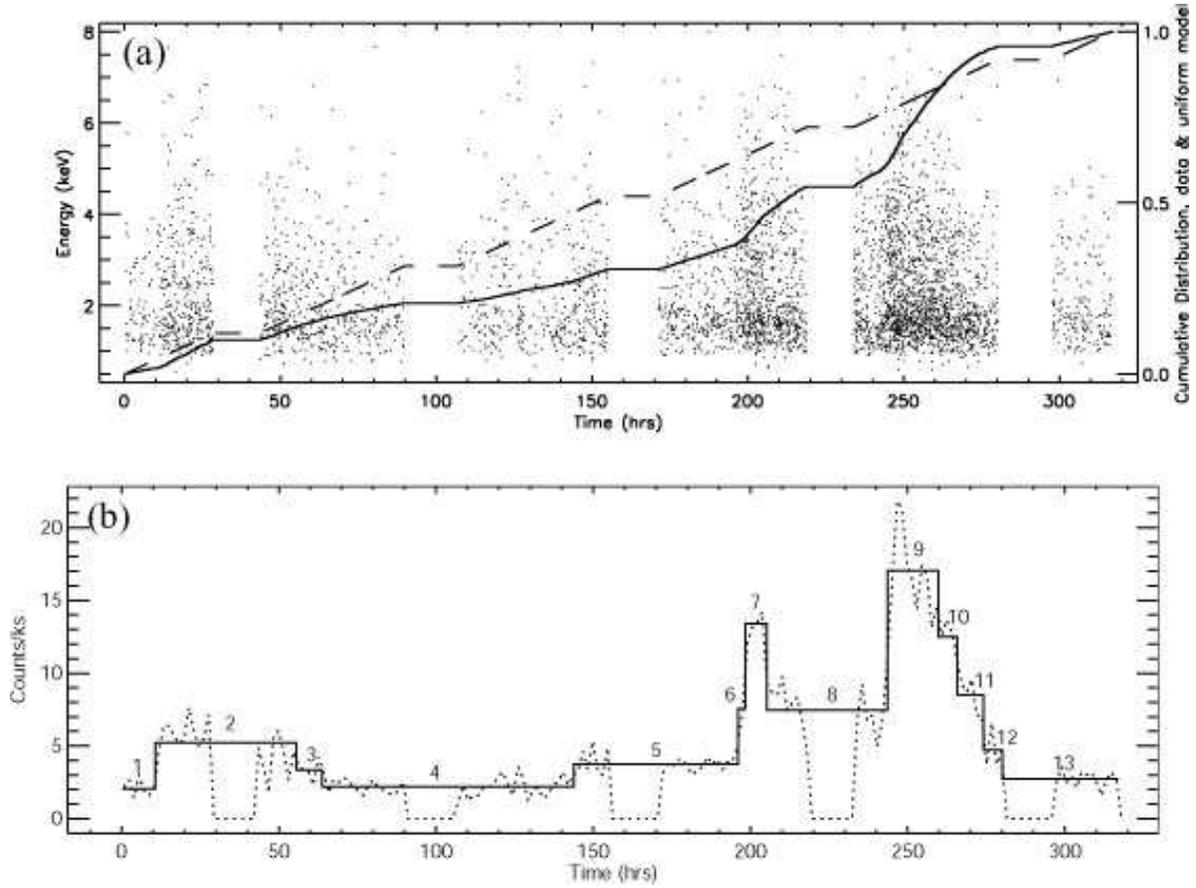} \hspace{0.05in}
\caption{Variability analysis of COUP source \#645. (a) Arrival
times as a function of energy in the total ($0.5-8.0$ keV) energy
band. The overplotted lines represent the cumulative distributions
of the data (solid) and uniform model (dashed) used to compute KS
variability test. (b) Bayesian Blocks segmentation (solid line) of
the X-ray lightcurve (dotted line) in the total energy band.
\label{var_analysis_fig}}
\end{figure}

First, the {\rm{\it acis\_extract}} package applies the
nonparametric one-sample Kolmogorov-Smirnov (KS) test to establish
whether variations are present above those expected from Poisson
noise associated with a constant source.  Here the cumulative
distribution of photon arrival times is compared to a simple model
for a constant source subject to the same gaps in observation
times due to orbit perigees (Figure \ref{var_analysis_fig}a). In
Table \ref{COUP_var_table} Column 4 reports the logarithm of test
significance: probability values $\log P_{{\rm KS}} \leq -3.0$ can
be considered almost definitely variable, while variability has
not been reliably detected when $\log P_{{\rm KS}} > -2.0$.
Log~$P_{{\rm KS}}$ has been truncated at $-4.0$ for strong sources
with high amplitude variability because the tail of the statistic
distribution is not well-defined.

Second, we employed the Bayesian Block (BB) parametric model of
source variability developed by \citet{Scargle98}. Here, the
lightcurve is segmented into a contiguous sequence of constant
count rates (Figure \ref{var_analysis_fig}b). The change points
between constant count rates are determined by an iterative
maximum likelihood procedure for a Poissonian process. We modeled
the ACIS frame mode data stream as binned data, with modifications
to cancel the COUP bad time intervals. The maximum likelihood
procedure corresponding to binned data was implemented in the IDL
language\footnote{Description of this procedure and code can be
found online at
\url{http://www.astro.psu.edu/users/gkosta/COUP/BBCODE/}.}. BB
constant count rates and boundaries were corrected for CCD readout
dead times and COUP bad time intervals, respectively. The prior
ratio parameter indicates the subjective preference we have for
the single-rate Poisson model over the dual-rate model prior to
analyzing the data. This quantity is practically useful for
suppressing spurious segments due to the statistical fluctuations.
We found that the overall shape of the BB light curves is
insensitive to large changes in the prior ratio, from 1 up to the
ratio of the length of data interval to the desired time
resolution for bright COUP sources ($\sim 100-200$), that is the
condition to stop oversegmenting \citep{Scargle98}. Thus the prior
ratio parameter was assigned the unbiased value of 1.

A more important parameter is the minimum number of counts allowed
in a segment, which stops the segmenting process; in many cases,
the number of segments varies with this value. After
experimentation, we chose a minimum of 5 counts per segment in
order to achieve maximal consistency between the BB and KS tests
for marginally variable sources. Specifically, if we define a
source to be BB variable if the number of segments ${\rm BBNum
\geq 2}$ and a source to be KS variable if $\log P_{{\rm KS}} <
-2.0$, then 1515 of 1616 COUP sources are consistently classified
as variable or non-variable by the two methods. With these
criteria for variability, 60\% (974 for KS or 973 for BB) of the
1616 COUP sources are variable. One consequence of combining the
unbiased prior ratio parameter with only 5 minimum counts per
segment is that strong sources often trigger isolated short
segments with a small count number. These spurious peaks are
easily identified by visual examination of the BB light curve.

Columns $5-9$ of Table \ref{COUP_var_table} report the results of
the BB variability analysis.  They give the number of segments
${\rm BBNum}$, the count rates of the lowest and highest segments,
and their $1\,\sigma$ uncertainties \citep[from][]{Gehrels86}
assuming Poisson statistics within the segment. The minimum count
rate might sometimes be viewed as the quiescent level between
flaring events. The ratio of maximum to minimum count rates may,
within errors, be viewed as a measure of variability amplitude. We
caution, however, that visual examination and individual analysis
of lightcurves is needed for a full understanding of COUP source
variability.

\section{Luminosities}\label{luminosities_section}

X-ray luminosities are provided here in three broad bands: $L_s$
in the soft $0.5-2.0$ keV band; $L_h$ in the hard $2.0-8.0$ keV
band; and $L_t$ in the total $0.5-8.0$ keV band.  Luminosities are
calculated from fluxes $F$ in these bands according to $L = 4 \pi
D^2 F$ assuming a distance $D=450$ pc to the Orion Nebula region.
For 1543 of the 1616 COUP sources, fluxes are obtained from the
thermal plasma spectral fits (\S \ref{spectral_analysis_section})
using {\rm{\it XSPEC's}} {\it flux} tool which integrates the
model spectrum over the desired band. Since the ARFs used with
{\rm{\it XSPEC}} incorporate instrumental effects, such as the
unextracted PSF fraction and absorption by hydrocarbon
contamination on the detector, no additional correction factors
are needed at this stage. Running {\rm{\it XSPEC's}} {\it flux}
tool with the fitted plasma energies and emission measures but
with zero absorption gives estimates of the intrinsic source
emission prior to interstellar absorption.  We call these
absorption-`corrected' luminosities $L_{s,c}$, $L_{h,c}$ and
$L_{t,c}$.  Formal $\sqrt{N}$ statistical uncertainties on the
luminosity can be readily obtained from ${\rm NetCts}$ and ${\rm
HR1}$ values in Table \ref{COUP_pho_table}. For most sources,
these statistical uncertainties to the luminosities are only about
$\pm 0.1$ or less in $\log L$.

The scientific reliability of these luminosities, however, is
usually considerably lower than the formal statistical uncertainty
for several reasons.  First, 60\% of the sources exhibit temporal
variability, often by factors of $0.3-1.3$ in $\log L_t$ (\S
\ref{variability_analysis_section}). Second, the measured value of
$L_s$ is often only a small fraction of the emitted soft
luminosity due to interstellar absorption.  There are $>300$
sources with fitted $\log N_H \geq 22.5$ cm$^{-2}$; for such
heavily absorbed sources, the soft plasma component and $L_s$ is
essentially unknown from spectral fits. The average lightly
absorbed source has $L_s \simeq 2 L_h$, so that the majority of
the luminosity is probably missed in the heavily absorbed sources.
The absorption-corrected value $\log L_{s,c}$ is so uncertain that
we do not provide it for scientific analysis.

\begin{figure}
\centering
\includegraphics[width=4.0in]{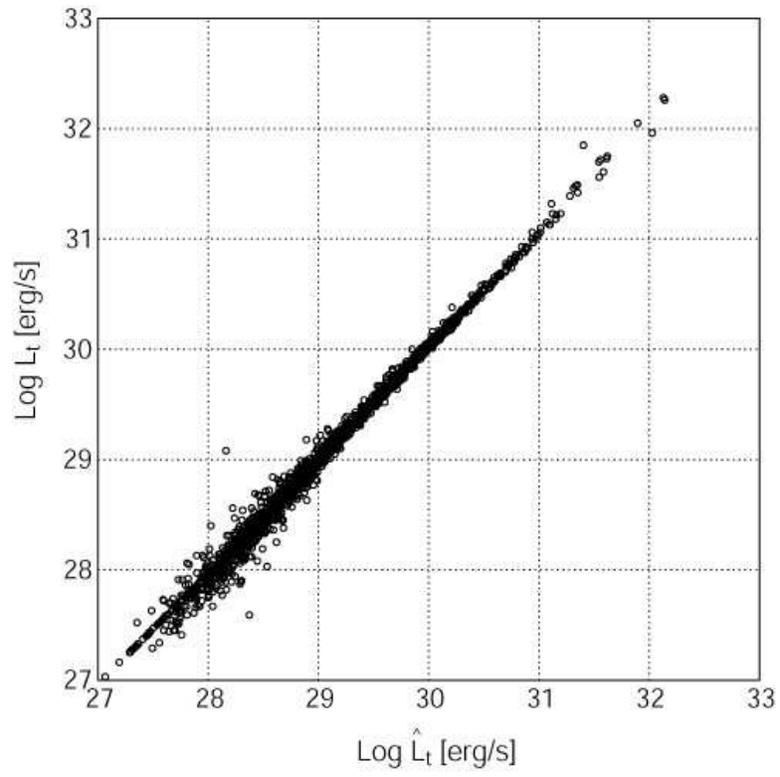}
\caption{Scatter plot of total band luminosities $L_{t}$ inferred
from {\rm{\it XSPEC}} spectral fitting with the approximate
luminosity $\hat{L_t}$ derived from photometric data.
\label{Lt_vs_Lhat_fig}}
\end{figure}

For the 73 faintest sources without any {\rm{\it XSPEC}} spectral
fit, we estimate the total band luminosities $\hat{L_t} \varpropto
{\rm IncFl \times MedE}$, where the incident flux ${\rm IncFl}$
and source's median energy ${\rm MedE}$ are given in Table
\ref{COUP_pho_table}. The basis for this approximation is the
strong correlation between the $\log L_t$ values derived from the
{\rm{\it XSPEC}} spectral fits with the approximate $\hat{L_t}$
values shown in Figure \ref{Lt_vs_Lhat_fig}. These values
generally differ from each other by no more than 0.2 in $\log L$.

Columns $3-7$ of Table \ref{COUP_lum_table} list $\log L_s$, $\log
L_h$, $\log L_{h,c}$, $\log L_t$ and $\log L_{t,c}$.

\section{Stellar Counterparts}\label{stellar_counterparts_section}
COUP source positions were compared with source positions from two
optical catalogs, two near-infrared ($JHK_s$ bands) catalogs, and
two thermal-infrared ($L$ band) catalogs. These catalogs differ in
sensitivity and fields of view.  In the optical band, we use the
catalog of \citet[][henceforth H97]{Hillenbrand97} and additional
stars found by \citet[][henceforth H02]{Herbst02}. Both of these
surveys cover the full COUP field. In the $JHK_s$ bands, we use a
merged catalog developed by McCaughrean et al.\ (in preparation;
henceforth McC04) for the inner, most sensitive quarter of the
COUP field and the 2 Micron All-Sky Survey \citep[][henceforth
2MASS]{Cutri03} for the outer portions of the field. Many of the
2MASS nondetections (i.e., 2MASS sources not detected in the COUP
image) can be attributed to spurious sources in the 2MASS catalog
which lie along bright emission-line filaments of the nebula (such
as the Orion Bar). In the $L$-band, we use the catalog of
\citet[][henceforth LMLA]{Lada04} for a small inner region and the
catalog of \citet[][henceforth MLLA]{Muench02} for additional
coverage around the center of the COUP field.

The $JHK_s$ catalog provided by McC04 is particularly valuable for
studies in the inner $7\arcmin \times 7\arcmin$\ of the COUP
field. It starts with a deep imaging survey of the inner Trapezium
Cluster made using the ISAAC near-infrared camera on UT1 of the
ESO Very Large Telescope which reaches 5$\sigma$ point source
limiting magnitudes of approximately 22, 21, and 20, at $J_s$,
$H$, and $K_s$, respectively. The seeing is roughly 0.5--0.6
arcsec FWHM throughout and thus a good match to the ACIS PSF in
this inner crowded region.  Its 1204 point sources are
astrometrically tied to the 2MASS reference frame with
0.15\arcsec\/ root-mean-squared accuracy. As the VLT data are
saturated for sources brighter than $\sim$13 magnitudes in all
filters, magnitudes for brighter stars were from other wide-field
ONC catalogs and high spatial resolution studies of the inner core
region as necessary \citep[][and
2MASS]{McCaughrean94,Hillenbrand97,Petr98,Simon99,Lucas00,Luhman00,Muench02,Schertl03}.
Care was taken to place all of the photometric data on the 2MASS
$JHK_s$ color system as far as possible, resulting in a
homogeneous merged catalog covering the brightest OB stars down to
candidate 3--5\,M$_{\rm Jup}$ proto-brown dwarfs in this inner
region.

For each catalog, we performed automated cross-correlations between
COUP and catalog source positions within a search radius of $1\arcsec$
for COUP source within $\simeq 3.5\arcmin$ of the field center, and within a
search radius of $2\arcsec$ in the outer regions of the field where the
$Chandra$ point spread function deteriorates. The results of this
search are given in Tables \ref{COUP_opt_table} and
\ref{COUP_nir_table}.

\begin{figure}
\centering
\includegraphics[angle=0.,width=7.0in]{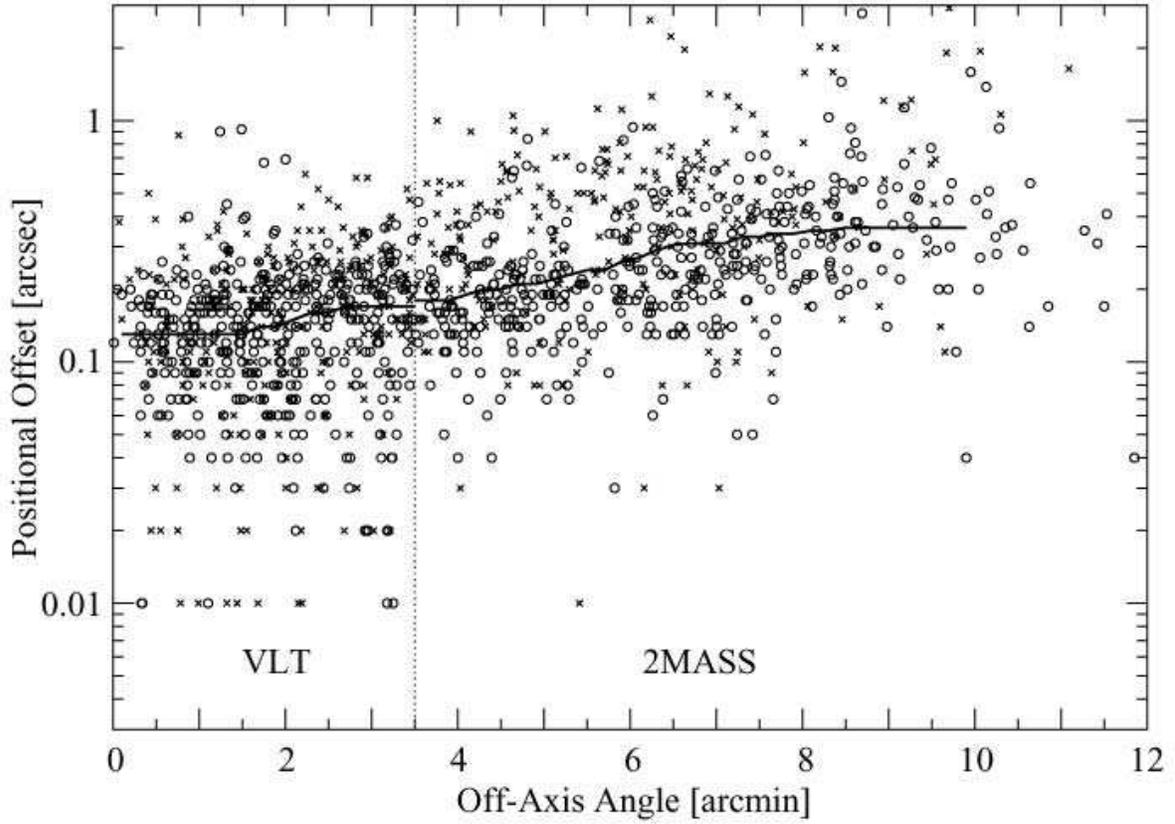}
\caption{Separations between COUP X-ray sources and IR
counterparts. For each individual source symbols present
separations between COUP and VLT catalog within the inner $<
3.5\arcmin$ region, and COUP and 2MASS catalog within the outer $>
3.5\arcmin$ region. X indicate weak X-ray sources with $< 200$
counts, circles indicate bright X-ray sources with $\geq 200$
counts. Solid lines present the running medians of separations.
\label{absolute_astrometry_fig}}
\end{figure}

To evaluate the merits of this automatic procedure, we performed a
careful visual examination of each COUP source. In $\simeq$98\% of the
cases, the identifications are clear and unambiguous. Median offsets
between COUP sources and near-infrared stars are only 0.15\arcsec, and between
COUP and optical stars are only 0.24\arcsec\footnote{Hillenbrand's
positions have been shifted by +0.78\arcsec\ in right ascension and
-0.38\arcsec\ in declination from the original publication.
\label{system_shift_footnote}} indicating superb astrometric positions
in the COUP field. Figure \ref{absolute_astrometry_fig} shows
individual offsets as a function of off-axis angle; median offsets are
better than $0.2\arcsec$ at the central part of the field and
$0.4\arcsec$ at larger off-axis angles.  Note that the (unexplained)
deterioration of several arcseconds in off-axis positions reported by
\citet{Feigelson02a} in earlier Orion ACIS data is not present in COUP
positions.

But in a few dozen cases, stellar identifications are complicated
due to closely spaced stars, many of which are probably physical
binaries.  In some cases, a single COUP source spans double NIR
and/or optical systems (see notes on individual sources in
electronic versions of Tables \ref{COUP_opt_table} and
\ref{COUP_nir_table}), while other cases have double COUP sources
with a single optical/infrared star. In the former case, the
closest optical/IR source is tentatively assigned to a COUP
source; in the latter case, either the same optical/infrared
counterpart is assigned to two COUP sources (if the 90\% PSF
contours both encompass the optical/IR position) or an
optical/infrared counterpart is tentatively assigned to the
closest COUP source. These cases, and other situations where
stellar identifications may be insecure, will be discussed
separately in Getman et al.\ (in preparation).  The listings in
Tables \ref{COUP_opt_table} and \ref{COUP_nir_table} are thus
probable, and not necessarily confirmed, physical associations
between COUP sources and previously identified stars.

In the inner region covered by the deep VLT survey, there are a
total of 883 COUP sources detected, 738 of which have stellar
counterparts in the McC04 merged catalog. However, that catalog
contains a total of 1204 stellar sources in the same region,
indicating that the COUP detection rate of cluster members is
$\sim$61\%. The great majority of these non-recovered sources are
faint, low-mass members of the cluster, most of which lie in the
brown dwarf regime. A careful discussion of the reasons for this
fall-off in COUP detections at the lowest masses and their
consequences for the X-ray activity of brown dwarfs will be given
in a future paper.

We find that 273 (17\% of 1616) COUP sources are not identified
with any of the known optical/IR sources. Nearly 90\% of these are
hard sources with ${\rm MedE > 2}$ keV, which is a clear indicator
of heavy absorption, and they constitute nearly 40\% of all hard
COUP sources. However, by itself, this does not determine which of
these unidentified sources are new PMS stars associated with the
Orion molecular cloud and which are background (primarily
extragalactic) sources.

For the probable optical stellar counterparts in Table
\ref{COUP_opt_table}, the star identifier is given in Column 3
where designations between 0 and 9999 are from H97 and 10000-11000
are from H02.  Column 4 gives the offset between the optical and
COUP positions.  The remaining 15 columns provide optical
measurements and inferred stellar properties based largely on the
comprehensive study of H97.  Columns $5-6$ give $V$ and $I$
magnitudes from H97. Spectral types in Column 7 are from H97
updated with the work of \citet{Luhman00} and \citet{Lucas01}.
Effective temperatures (Column 9) are based on spectral types
following \citet{Hillenbrand04}, while visual absorption $A_V$
(Column 8), bolometric luminosity $\log L_{\rm{bol}}$ (Column 10),
and radius $R$ (Column 11) are derived as in H97. Columns 12 and
13 are the mass $M$ and age $\log t$ recalculated by us using the
PMS evolutionary tracks of \citet{Siess00}.  Users should be
cautious when using the intrinsic stellar properties as they are
less accurate than the precision given in the tables.  Errors
arise from intrinsic photometric variability, uncertainties in
spectral types and conversions to effective temperatures and
bolometric luminosities, and, for the mass and age estimates,
uncertainties in theoretical evolutionary tracks
\citep[see][]{Hillenbrand04}.

Columns $14-15$ give two properties associated with circumstellar
material and accretion measured by \citet{Hillenbrand98}: the
$K$-band excess attributable to a hot circumstellar disk,
$\Delta(I-K)$, and the equivalent width of the Ca~${\rm{II}}$
infrared triplet lines with $\lambda = 8542\AA$, $EW({\rm{Ca}})$.
A negative value here represents an emission line.  Columns
$16-19$ give properties derived from $\sim 90$ epochs of
photometric monitoring by H02: the average $V$ magnitude $<V>$ and
its standard deviation $\Delta V$, the range in $V$ about this
mean, and the rotational period $P$ derived from periodicity
attributed to rotationally modulated starspots.

Individual near-infrared counterparts for COUP sources are listed
in Table \ref{COUP_nir_table} along with infrared properties.
Column 3 gives the source identifier obtained from the merged
catalog of McC04 for the inner region of the COUP field, and the
2MASS designation for the outer region.  The positional offsets
and $JHK_s$ magnitudes from the appropriate catalog appear in
columns $4-7$.  Column 8 gives a flag for McC04 sources indicating
the source of the photometry: 1 from the VLT; 2 and 3 from the NTT
and FLWO surveys, respectively, presented by \citet{Muench02}; 4
from the compilation of H97, and 5 from 2MASS or other sources.
Columns $9-10$ give flags for 2MASS sources. The values of the
photometry quality flag in column 9 are: A = signal-to-noise ratio
(SNR) $\geqslant10$; B = SNR $\geqslant7$; C = SNR $\geqslant5$; D
= low significance detection; E = PSF fitting poor; F = reliable
photometric errors not available; X = source detected but no valid
photometry is available.  The values of the confusion and
contamination flag in column 10 are: 0 = no problem; b = possible
multiple source; c = photometric confusion from nearby star; d =
diffraction spike confusion from nearby star; p = persistence
contamination from nearby star; and s = electronic stripe from
nearby star.

Columns $11-13$ of Table \ref{COUP_nir_table} refer to L band data.
The first column indicates whether the photometry is obtained from the
surveys of \citet[][designated M]{Muench02} or \citet[][designated
L]{Lada04}.  The second column gives the running source number from the
L band catalog.  The final column gives the L magnitude.

\section{Nondetections of Cluster Members}\label{nondetections_section}

Tables \ref{COUP_lim opt_table}$-$\ref{COUP_lim 2m_table} provide
lists of upper limits for the Hillenbrand (1997) $V<20$ optical catalog
and the 2MASS $K<15$ near-infrared catalog.  The first three columns in
the tables give object identification name or number and star position
from the associated catalog. The next three columns give information
from the COUP image: the upper limit on the number of counts at that
position, effective exposure time at the object position, and a
confusion flag if the object is influenced by nearby sources. The
remaining columns reproduce catalog optical and near-infrared
properties for the stars (see \S \ref{stellar_counterparts_section}).
We underscore the importance of the contamination and confusion flags
in the 2MASS dataset, as a considerable number of 2MASS cataloged sources
are small-scale bright spots in the diffuse nebular emission rather
than true stars.

The upper limits to the COUP counts for these sources were
calculated using {\rm {\it PWDetect}} and are therefore consistent
with the {\rm {\it PWDetect}} detection procedure. They indicate
the minimum source counts required, for a detection using {\rm
{\it PWDetect}} at a significance level of 5.0\,$\sigma$, i.e.,
the same used for detection. Our final source list contains some
sources with fewer counts than the upper limits derived by {\rm
{\it PWDetect}}. This can be due to: 1) the use of other detection
methods, which can, in some situations, be more sensitive than
{\rm {\it PWDetect}} (e.g., the eye is able to detect a faint
source close to a bright one more easily than a wavelet based
detection algorithm, especially where the PSF is azimuthally
distorted); 2) random fluctuations in the spatial distribution of
source events raising the S/N of a source above the detection
threshold. Such a source would be undetectable if it had a nominal
PSF.

For most stars, the upper limits are estimated analytically from
the source-free background map computed by {\rm {\it PWDetect}}.
In many cases where the undetected stars lie on PSF wings of a
detected source (as indicated by the flag in Tables \ref{COUP_lim
opt_table}$-$\ref{COUP_lim 2m_table}), the analytical method may
significantly overestimate upper limits. More stringent limits
were derived for such sources by adding test sources at a variety
of count levels at the upper limit position in the ACIS image. The
wavelet transforms of these simulated images were then examined to
determine the minimum count level (reported as the upper limit in
our tables) at which the simulated source becomes detectable
(i.e., produces a second peak next to that of the detected
source).

It is often desirable to estimate the luminosity upper limits and
we include a possible procedure below (these luminosity upper
limits will be included in a future COUP publication).  Derive an
upper limit to the count rate as ${\rm LimCt/Exp}$ using Tables
\ref{COUP_lim opt_table}$-$\ref{COUP_lim 2m_table}, calculate
$N_{\rm H}$ according to the formula $N_{\rm H}({\rm cm^{-2}})
\sim A_V {\rm (mag)} \times 1.6 \times 10^{21}$ \citep{Vuong03},
use {\rm PIMMS} to compute the unabsorbed flux from the count
rates assuming an optically thin plasma spectrum with $kT \sim 1$
keV. Although the mean temperature of the detected COUP sources is
somewhat higher than 1 keV, there is a trend of increasing
temperature with increasing X-ray luminosity, and since we can
expect the undetected objects to be more similar to the faintest
X-ray sources, 1 keV should be a reasonable value.

\section{Maps and Source Atlas}\label{atlas_section}

\begin{figure}
\centering
\includegraphics[angle=0.,width=5.0in]{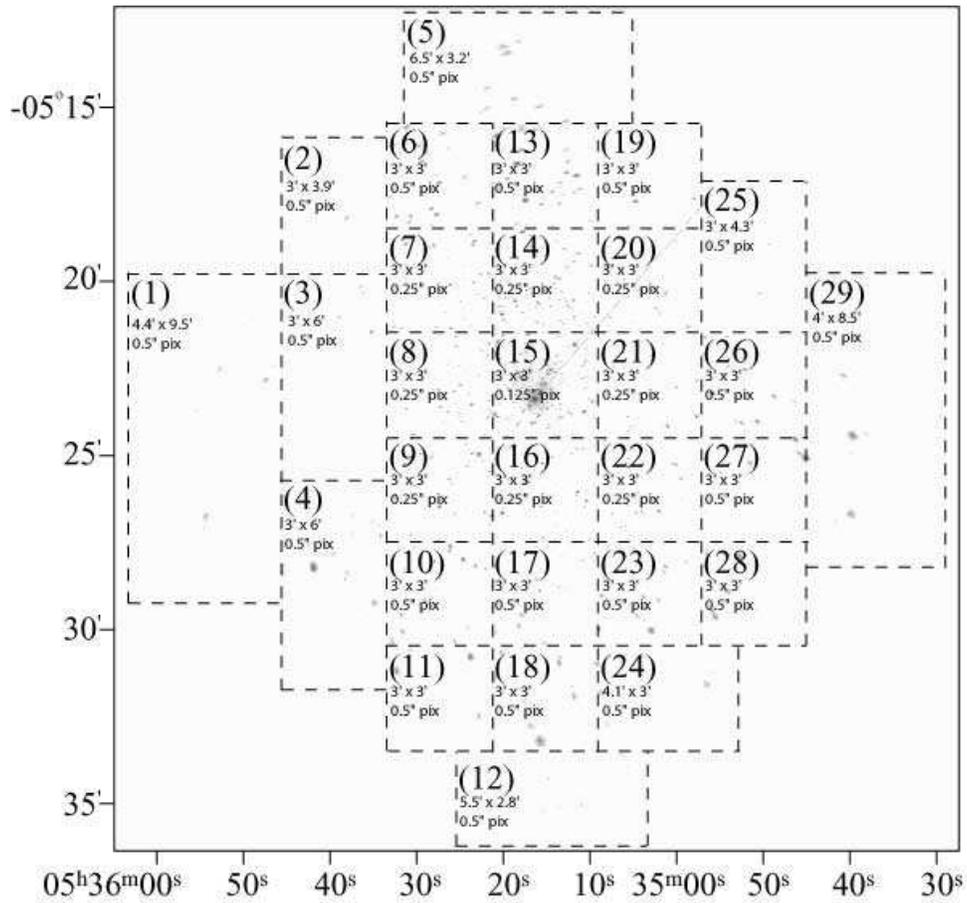}
\caption{Guide to the expanded views of the merged COUP ACIS-I
data. \label{guide_fig}}
\end{figure}

Figure \ref{expanded_view_fig} provides closeup maps covering the
entire merged COUP ACIS-I field with 1616 COUP sources labelled.
The field was divided into 29 panels with the innermost panel
having the highest resolution.  A guide to the map panels is
provided in Figure \ref{guide_fig}; this guide also gives the
pixel size for each panel. The plotted intensities are scaled to
the log of the photons in each pixel.

\begin{figure}
\centering
\includegraphics[angle=0.,width=6.0in]{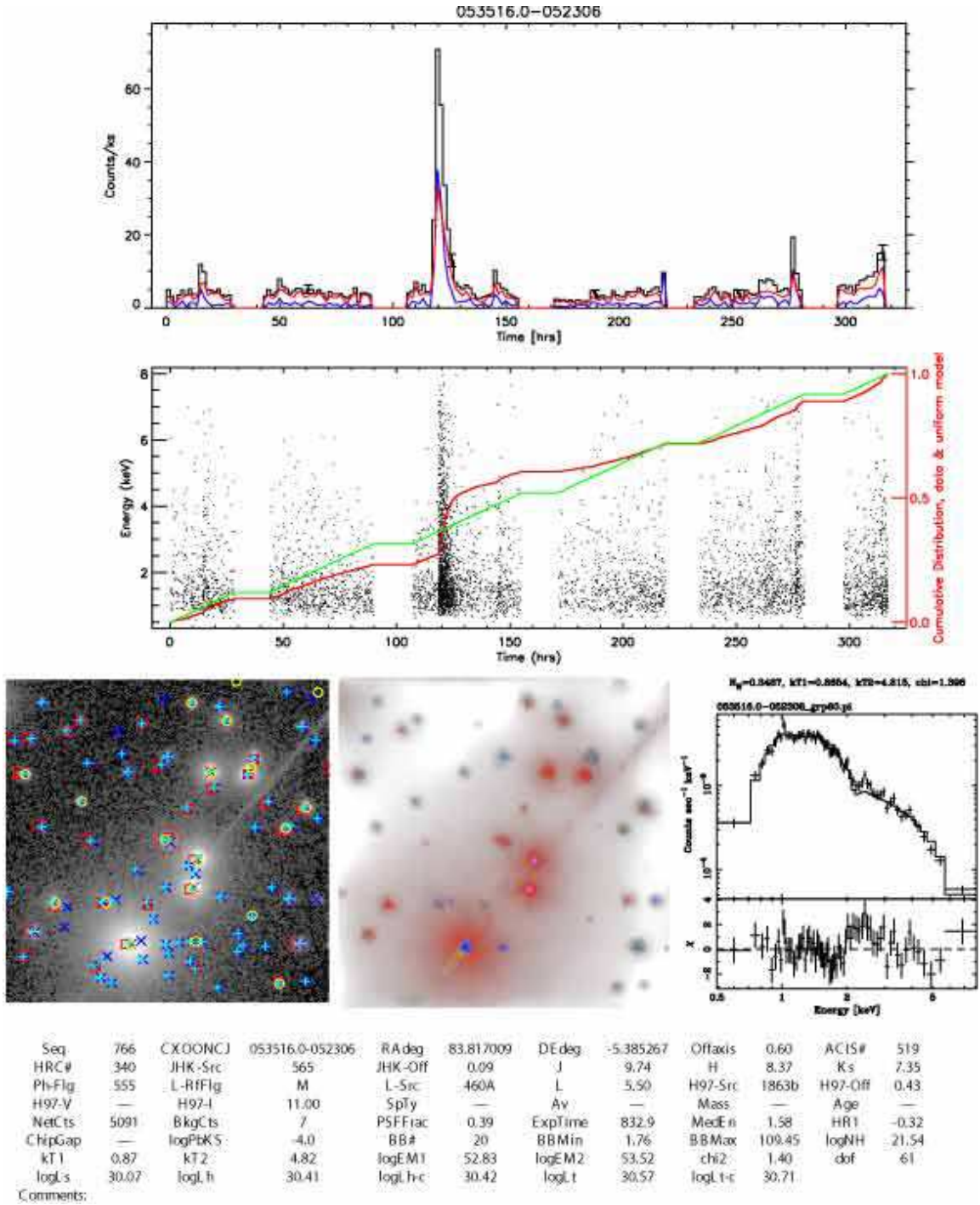}
\caption{Sample page from the COUP source atlas.
\label{sample_atlas_fig}}
\end{figure}

The {\rm{\it acis\_extract}} package also produces an ``atlas''
where both tabulated and graphical information on each source is
collected onto a single page.  Figure \ref{sample_atlas_fig} shows
a sample page.  The graphs display various projections of the
four-dimensional ACIS dataset -- photons as a function of RA, Dec,
energy, and time -- with ancillary plots associated with our
spatial, spectral, and variability analysis\footnote{The 1616
pages of the COUP atlas are
available in PDF format at\\
\url{http://www.astro.psu.edu/users/gkosta/COUP/DATA\_PRODUCTS\_08\_17\_04/ATLAS/.}}.

After the COUP source name, the upper panel of each atlas page
shows the lightcurve of extracted events without background
subtraction. In the few cases where the source lies near a gap in
the CCD array, it covers only the primary CCD. The abscissa is in
hours from the start of the observation and the ordinate is in
counts ks$^{-1}$.  The black histogram shows the total $0.5-8.0$
keV energy band, while the red lines show the soft ($0.5-2.0$ keV)
and the blue lines show the hard ($2.0-8.0$ keV) energy bands.
The light curve binning depends on source strength according to
the following scheme:  $\rm{SrcCnts} < 200$, bin width $= 6.3$
hours; $<500$, 3.2 hours; $<10000$, 1.6 hours; $<20000$, 1.1
hours; $<40000$, 47.6 minutes; and $\geq 40000$, 23.8 minutes.

The second panel presents the photon arrival times as a function
of energy in the total energy band.  The overplotted lines
represent the cumulative distributions of data (red) and uniform
model (green), used to compute the K-S variability test (\S
\ref{variability_analysis_section}).

The bottom left and center panels show the raw and adaptively
smoothed images around the source. The images subtend $50.5\arcsec
\times 50.5\arcsec$ size with $0.25\arcsec$-pixel bins oriented
with north at the top and east to the left. Brightness is scaled
to the logarithm of counts in each pixel; isolated individual
events are always visible in the raw image. The raw image shows a
green polygon representing the {\rm{\it acis\_extract}} source
extraction region and a red circle representing the source mask
used in construction of the Penn State background map ($1.1 \times
97\%$ PSF region). Small colored symbols indicate the location of
IR and optical sources: cyan $+$ are from VLT $JHK$ catalog,
yellow circles are from the 2MASS $JHK_s$ catalog, blue $\times$
are from the MLLA $JHKL$ catalog, and red boxes are from the
optical catalog of Hillenbrand (1997). The adaptive smoothing is
performed using the CIAO tool {\rm{\it csmooth}} at the
$2.5\sigma$ level. The adaptively smoothed image in the center
panel is a "true-color"\footnote{Red, green, and blue images are
scaled using a color model (from the TARA package {\url
http://www.astro.psu.edu/xray/docs/TARA/}) that preserves the
``hue'' of the data (ratios among red, green, and blue signals),
even for regions where the brightness is saturated. The color
white and other shades of grey appear only where the red, green,
and blue signals are approximately equal. A similar technique has
been described by \citet{Lupton03}.} X-ray image where red
represents the $0.5-1.7$ keV band counts, green the $1.7-2.8$ keV
counts, and blue the $2.8-8.0$ keV counts.

The bottom right panel of each atlas page shows the extracted
source spectrum after background subtraction and the chosen model
from the {\rm{\it XSPEC}} spectral fitting procedure (\S
\ref{spectral_analysis_section} and Table \ref{COUP_spe_table}).
The top plot shows the source spectrum after grouping as separated
lines with error bars and a continuous step-function representing
the fitted model. The abscissa is in units of log (keV) and the
ordinate is in units of log (cts s$^{-1}$ keV$^{-1}$). The bottom
plot shows residuals in units of contribution to the $\chi^2$
statistic.

Finally, below the figures on each page of the COUP atlas we
collect many of the X-ray positional, photometric, spectral,
variability, and stellar counterpart quantities given in Tables
\ref{COUP_src_table} - \ref{COUP_lum_table}.

\section{Summary}\label{summary_section}

\begin{figure}
\centering
\begin{minipage}[t]{1.0\textwidth}
  \centering
  \includegraphics[angle=0.,scale=0.8]{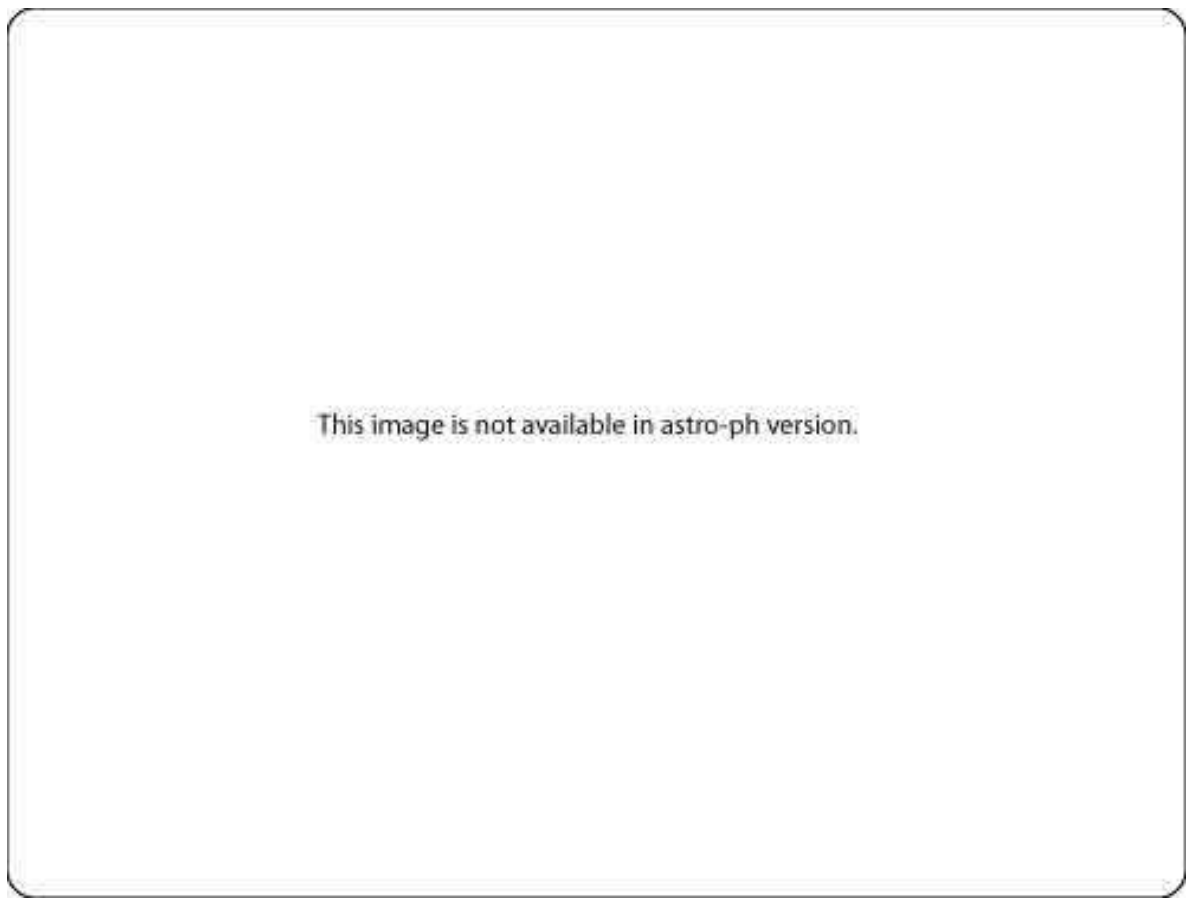} \hspace{0.05in}
\caption{Expanded views of the merged COUP ACIS-I data with
sources indicated.  Panel numbers are indicated in parentheses in
the upper left corner of each panel and refer back to
Figure~\ref{guide_fig}. \label{expanded_view_fig}}
\end{minipage}
\end{figure}

This paper presents the observations, data analysis methodology, and
tabulated results for the $Chandra$ Orion Ultradeep Project.  We
describe techniques used in the COUP data reduction, derivation of
time-averaged source properties, and production of the source catalog.
The results of this study are the tables of the 1616 COUP source
properties (Tables \ref{COUP_src_table}, \ref{COUP_pho_table},
\ref{COUP_spe_table}, \ref{COUP_var_table}, \ref{COUP_lum_table},
\ref{COUP_opt_table}, and \ref{COUP_nir_table}), tables of undetected
sources (Tables \ref{COUP_lim opt_table} and \ref{COUP_lim 2m_table}),
map of the field (Figure \ref{expanded_view_fig}), and the 1616 page
atlas detailing the analysis and findings for each source.

Our data reduction procedure seeks to use the most well-developed
methods available for the identification and extraction of point
sources in $Chandra$ ACIS images.  The source detection procedure
is based on two wavelet-based search algorithms, run on datasets
constructed in several bands, supplemented by visual inspection of
the image. The dataset is optimized for maximum reduction of
background and maximum spatial resolution with subpixel event
repositioning. We identify 1616 X-ray sources in the $\sim
17\arcmin \times 17\arcmin$ ACIS-I field.

For each source, we perform data extraction, pile-up analysis,
spectral analysis, variability analysis, and broad-band luminosity
determinations using the sophisticated semi-automated IDL-based
{\rm{\it acis\_extract}} package.  The algorithms used in {\rm{\it
acis\_extract}} successfully address many difficulties related to
the COUP observation such as the position-dependent point spread
function, multiple exposures with different pointings and time
gaps, source crowding, piled-up sources, nonuniform local
background and exposure map, and so on. Our treatment of photon
pile-up for strong sources using annular extraction regions is
first used here in a formal procedure. The {\rm{\it
acis\_extract}} package, which is available to the public, can be
used to create point source catalogs for many $Chandra$ studies.

These images, atlases and tables serve as the foundation for
various studies of the COUP observations. These will include
investigation of X-ray emission from different classes of pre-main
sequence stars: OB, intermediate-mass, T Tauri and brown dwarfs;
deeply embedded stars which are often younger protostars; and
binary and multiple star systems.  We will also investigate the
astrophysical issues such as the magnetic field generation and
geometries in pre-main sequence systems, plasma energetics in
powerful magnetic reconnection flaring events, and X-ray
ionization of circumstellar disks and cloud material. Altogether,
the COUP data products presented here offer the most detailed and
comprehensive information to date concerning the X-ray emission
and magnetic activity in pre-main sequence stars.

We thank George Chartas (Penn State) for providing the ARF
correction tool, $\rm {\it xpsf}$, and acknowledge the unusual
efforts of the {\it Chandra} Mission Operations group in
scheduling the COUP observations. We also thank the referee, M.
Muno, for his time and many useful comments that improved this
work. COUP is supported by {\it Chandra} guest observer grant SAO
GO3-4009A (E.\ D.\ Feigelson, PI). This work was also supported by
the ACIS Team contract NAS8-38252, HRC Team contract NAS8-39073.
EF, FD, GM and SS acknowledge financial contribution from Italian
MIUR (COFIN-PRIN 2001) and from INAF. MJM and G. Meeus acknowledge
support from the European Commission Research Training Network
``The Formation and Evolution of Young Stellar Clusters''
(HPRN-CT-2000-00155). This publication makes use of data products
from the Two Micron All Sky Survey, which is a joint project of
the University of Massachusetts and the Infrared Processing and
Analysis Center/California Institute of Technology, funded by the
National Aeronautics and Space Administration and the National
Science Foundation.

\bibliography{aj-jour}

\begin{figure}
\centering
\includegraphics[angle=0.,width=6.5in]{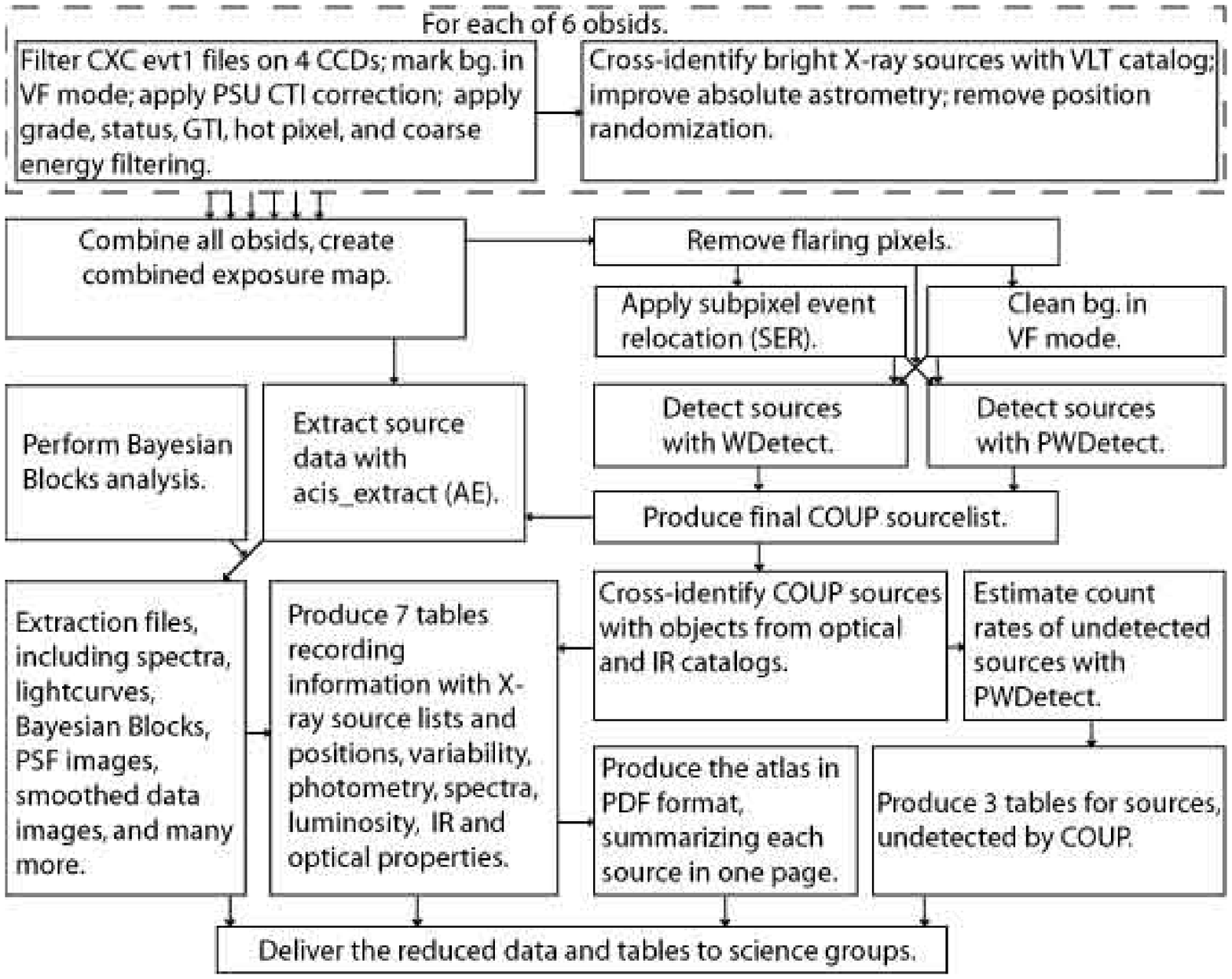}
\caption{Diagram of the COUP data reduction and catalog procedure.
\label{diagram_fig}}
\end{figure}
\clearpage
\newpage

\clearpage
\newpage

\begin{figure}
\centering
\includegraphics[width=1\textwidth]{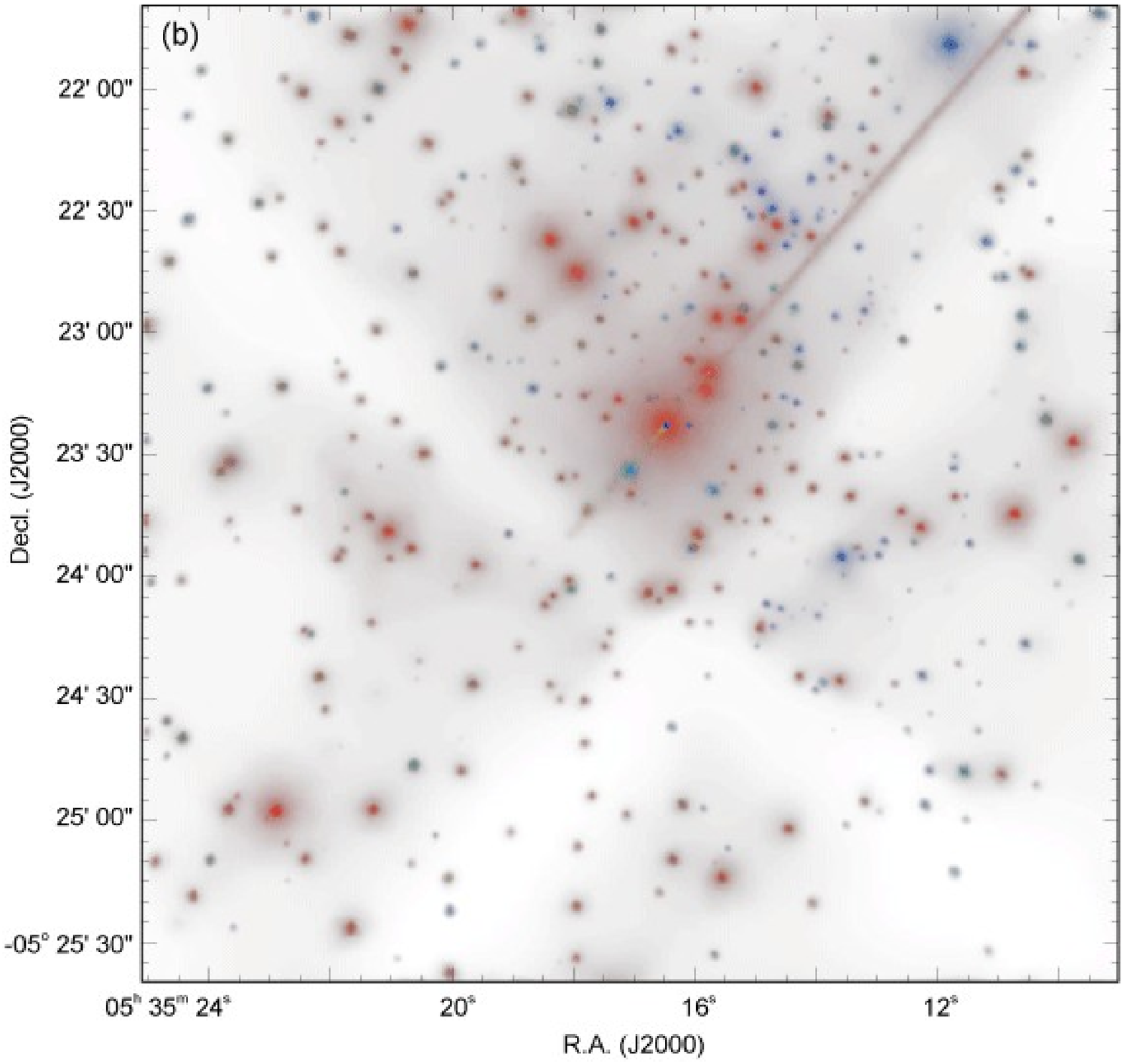}
\end{figure}

\clearpage
\newpage

\clearpage
\newpage

\clearpage
\newpage

\clearpage
\newpage

\clearpage
\newpage

\clearpage
\newpage

\clearpage
\newpage

\clearpage
\newpage

\clearpage
\newpage

\clearpage
\newpage

\clearpage
\newpage

\begin{deluxetable}{ccrccc}
\centering \tabletypesize{\small} \tablewidth{0pt}
\tablecolumns{6}

\tablecaption{COUP Observations \label{observation_journal_table}}

\tablehead{

\colhead{ObsID} & \colhead{Start Time} & \colhead{Exposure} &
\colhead{Right Ascension} & \colhead{Declination}
& \colhead{Roll Angle} \\

&& (ks) & \multicolumn{2}{c}{(degrees J2000)} & degrees }

\startdata
4395 & 2003 Jan 08 20:58:18.8 &  99.9 & 83.8210052 & -5.394400 & 310.90882 \\
3744 & 2003 Jan 10 16:17:38.7 & 164.2 & 83.8210086 & -5.394406 & 310.90882 \\
4373 & 2003 Jan 13 07:34:43.5 & 171.5 & 83.8210098 & -5.394403 & 310.90882 \\
4374 & 2003 Jan 16 00:00:37.4 & 168.9 & 83.8210079 & -5.394404 & 310.90882 \\
4396 & 2003 Jan 18 14:34:48.3 & 164.5 & 83.8210094 & -5.394403 & 310.90882 \\
3498 & 2003 Jan 21 06:10:27.7 &  69.1 & 83.8210025 & -5.394402 & 310.90882 \\

\enddata

\tablecomments{ ObsID values are from the $Chandra$ Observation
Catalog.  Start Times are in UT.  Exposure times are the sum of
Good Time Intervals (GTIs) for the CCD at the telescope aim point
(CCD3) minus 1.3\% to account for CCD readouts. The aimpoints and
roll angles are obtained from the satellite aspect solution before
astrometric correction was applied.}

\end{deluxetable}

\clearpage
\newpage

\begin{deluxetable}{rrrrrrrrcccrrrrr}
\centering \rotate \tabletypesize{\scriptsize} \tablewidth{0pt}
\tablecolumns{16}

\tablecaption{COUP X-ray Source Locations\label{COUP_src_table}}

\tablehead{

\multicolumn{2}{c}{Source} && \multicolumn{4}{c}{Position} &&
\multicolumn{5}{c}{Detection} && \multicolumn{2}{c}{Previous} \\
\cline{1-2} \cline{4-7} \cline{9-13} \cline{15-16}

\colhead{Seq} & \colhead{CXOONCJ} && \colhead{R.A.} &
\colhead{Decl.} & \colhead{Err} & \colhead{$\theta$} &&
\colhead{Flag} & \colhead{Flag} & \colhead{Flag} &
\colhead{Signif}
& \colhead{Signif} && \colhead{ACIS} & \colhead{HRC} \\

\colhead{\#} &&& \multicolumn{2}{c}{(degrees J2000)} &
\colhead{\arcsec} & \colhead{\arcmin}&& \colhead{1} & \colhead{2}
& \colhead{3} & \colhead{A\_E} &
\colhead{PWD} && \colhead{\#} & \colhead{\#} \\

\colhead{(1)} & \colhead{(2)} && \colhead{(3)} & \colhead{(4)} &
\colhead{(5)} & \colhead{(6)} && \colhead{(7)} & \colhead{(8)} &
\colhead{(9)} & \colhead{(10)} & \colhead{(11)} && \colhead{(12)}
& \colhead{(13)}  }

\startdata

   1 &   053429.4-052337 &&  83.622730 & -5.393730 & 0.16 & 11.85 && 0d000x & 11 & 110 &  23.9 &  36.0   && \nodata & \nodata \\
   2 &   053429.5-052354 &&  83.623040 & -5.398490 & 0.11 & 11.83 && 0d000x & 11 & 110 &  41.0 &  94.4   && \nodata &  10     \\
   3 &   053433.9-052211 &&  83.641473 & -5.369767 & 0.59 & 10.83 && 00000x & 10 & 100 &   0.6 & \nodata && \nodata & \nodata \\
   4 &   053434.9-052507 &&  83.645653 & -5.418817 & 0.38 & 10.58 && 00000x & 10 & 100 &   0.5 & \nodata && \nodata & \nodata \\
   5 &   053438.2-052338 &&  83.659339 & -5.393952 & 0.35 &  9.66 && 000000 & 11 & 110 &   4.0 &   7.0   && \nodata & \nodata \\
   6 &   053438.2-052423 &&  83.659281 & -5.406534 & 0.09 &  9.69 && 0000w0 & 11 & 110 &  28.8 &  77.3   &&    2    & \nodata \\
   7 &   053439.7-052425 &&  83.665798 & -5.407138 & 0.01 &  9.30 && 00p000 & 11 & 110 & 198.6 & 452.7   &&    4    &  18     \\
   8 &   053439.8-052456 &&  83.665915 & -5.415641 & 0.12 &  9.35 && 0000w0 & 11 & 110 &  24.3 &  54.9   &&    3    & \nodata \\
   9 &   053439.9-052642 &&  83.666336 & -5.445000 & 0.04 &  9.73 && 0d000x & 11 & 110 & 123.3 & 239.6   &&    6    &  20     \\
  10 &   053440.0-052604 &&  83.666807 & -5.434478 & 0.30 &  9.52 && 0d0000 & 11 & 110 &   4.0 &  11.7   && \nodata & \nodata \\
  11 &   053440.8-052242 &&  83.670374 & -5.378408 & 0.04 &  9.05 && 00000x & 11 & 110 &  72.7 & 160.8   &&    9    &  21     \\
  12 &   053440.8-052639 &&  83.670220 & -5.444250 & 0.18 &  9.49 && 0d000x & 10 & 002 &  15.4 & \nodata &&    7    & \nodata \\
  13 &   053440.9-052559 &&  83.670779 & -5.433305 & 0.33 &  9.27 && 0d0000 & 11 & 110 &   1.8 &   6.7   &&    8    & \nodata \\
  14 &   053441.6-052357 &&  83.673450 & -5.399378 & 0.24 &  8.82 && 0000wx & 11 & 110 &   7.7 &  18.9   && \nodata & \nodata \\
  15 &   053441.7-052652 &&  83.673936 & -5.448017 & 0.12 &  9.36 && 0d000x & 11 & 110 &  25.3 &  64.9   &&   10    & \nodata \\

\enddata
\tablecomments{See \S \ref{source_list_section} for description of
the columns.  The full table of 1616 COUP sources is available in
the electronic edition of the Journal.}
\end{deluxetable}

\clearpage
\newpage

\begin{deluxetable}{crr}
\centering \tabletypesize{\small} \tablewidth{0pt}
\tablecolumns{3} \tablecaption{Tentative COUP Sources
\label{tentative_table}}

\tablehead{\colhead{CXOONCJ} & \colhead{R.A.} & \colhead{Decl.}\\
& \multicolumn{2}{c}{(degrees J2000)}}

\startdata

     053503.8-051758  & 83.765885   & -5.299651\\
     053503.9-052253  & 83.766622   & -5.381438\\
     053504.0-052809  & 83.766730   & -5.469410\\
     053504.5-051908  & 83.768862   & -5.318981\\
     053505.7-052616  & 83.774150   & -5.437980\\
     053508.6-052022  & 83.785996   & -5.339450\\
     053508.8-052015  & 83.786847   & -5.337515\\
     053509.8-052143  & 83.791170   & -5.362130\\
     053510.4-052430  & 83.793640   & -5.408600\\
     053510.7-051940  & 83.794591   & -5.327879\\
     053512.8-052156  & 83.803440   & -5.365780\\
     053513.2-052322  & 83.805251   & -5.389650\\
     053513.5-052209  & 83.806492   & -5.369288\\
     053515.7-052639  & 83.815740   & -5.444270\\
     053528.1-051456  & 83.867420   & -5.249130\\
     053528.1-051958  & 83.867327   & -5.333052

\enddata

\end{deluxetable}

\clearpage
\newpage

\begin{deluxetable}{rrrrrrrrrrrrrrrrrrrr}
\centering \rotate \tabletypesize{\scriptsize} \tablewidth{0pt}
\tablecolumns{20}

\tablecaption{COUP Source X-ray Photometry \label{COUP_pho_table}}

\tablehead{

\multicolumn{2}{c}{Source} && \multicolumn{5}{c}{Extraction} &&
\multicolumn{3}{c}{Flux} && \multicolumn{7}{c}{Hardness} \\
\cline{1-2} \cline{4-8} \cline{10-12} \cline{14-20} \\

\colhead{Seq} & \colhead{CXOONCJ} && \colhead{Src} & \colhead{Bkg}
& \colhead{Net} & \colhead{Area} & \colhead{PSF} && \colhead{Exp}
& \colhead{Ct-Fl} & \colhead{IncFl} && \colhead{Med E} &
\colhead{HR1} & \colhead{$\Delta {\rm HR1}$}
& \colhead{HR2} & \colhead{$\Delta {\rm HR2}$}& \colhead{HR3} & \colhead{$\Delta {\rm HR3}$} \\

\colhead{\#} &&& \colhead{Cts} & \colhead{Cts} & \colhead{Cts} &
\colhead{pix} & \colhead{Frac}&& \colhead{ks} & \colhead{(a)}
& \colhead{(b)} && \colhead{keV} &&&&&& \\

\colhead{(1)} & \colhead{(2)} && \colhead{(3)} & \colhead{(4)} &
\colhead{(5)} & \colhead{(6)} & \colhead{(7)} && \colhead{(8)} &
\colhead{(9)} & \colhead{(10)} && \colhead{(11)} & \colhead{(12)}
& \colhead{(13)} &  \colhead{(14)} & \colhead{(15)} &
\colhead{(16)} & \colhead{(17)}  }

\startdata

   1  &  053429.4-052337  &&    883~ &  89  &   793~ &  664 & 0.50 && 500.4 & 0.283 & -4.897 && 1.28~ &  -0.59  &   $_{-0.04}^{+0.04}$  & -0.54 & $_{-0.03}^{+0.04}$ & -0.34   &  $_{-0.08}^{+0.08}$   \\
   2  &  053429.5-052354  &&   1884~ &  53  &  1830~ & 1245 & 0.68 && 157.4 & 0.089 & -4.308 && 1.19~ &  -0.69  &   $_{-0.02}^{+0.02}$  & -0.67 & $_{-0.02}^{+0.02}$ & -0.28   & $_{-0.05}^{+0.05}$    \\
   3  &  053433.9-052211  &&    148~ & 132  &    15~ & 1541 & 0.89 && 382.0 & 0.216 & -6.737 && 0.88~ & -1.00   & $_{-...}^{+0.06}$ & -1.00 & $_{-...}^{+0.40}$ &  \nodata   & \nodata    \\
   4  &  053434.9-052507  &&    327~ & 310  &    16~ & 1678 & 0.89 && 527.0 & 0.298 & -6.986 && 5.02~ & 1.00 & $_{-0.00}^{+...}$ & \nodata & \nodata &  1.00   & $_{-0.00}^{+...}$    \\
   5  &  053438.2-052338  &&    139~ &  70  &    68~ &  396 & 0.69 && 740.9 & 0.419 & -6.292 && 4.69~ &   0.75  &   $_{-0.20}^{+0.19}$  & -0.22 & $_{-0.88}^{+0.45}$ &  0.81   & $_{-0.16}^{+0.17}$    \\
   6  &  053438.2-052423  &&   2260~ & 372  &  1887~ &  994 & 0.90 && 740.9 & 0.419 & -4.974 && 1.26~ &  -0.66  &   $_{-0.03}^{+0.04}$  & -0.61 & $_{-0.03}^{+0.03}$ & -0.46   & $_{-0.11}^{+0.11}$    \\
   7  &  053439.7-052425  &&  42315~ & 362  & 41952~ &  846 & 0.89 && 746.2 & 0.422 & -3.630 && 1.20~ &  -0.72  &   $_{-0.00}^{+0.00}$  & -0.67 & $_{-0.00}^{+0.00}$ & -0.37   & $_{-0.01}^{+0.01}$    \\
   8  &  053439.8-052456  &&   1349~ & 219  &  1129~ &  866 & 0.90 && 746.2 & 0.422 & -5.201 && 2.59~ &   0.25  &   $_{-0.04}^{+0.04}$  &  0.06 & $_{-0.05}^{+0.05}$ &  0.22   & $_{-0.04}^{+0.04}$    \\
   9  &  053439.9-052642  &&  16471~ & 199  & 16271~ & 1318 & 0.99 && 385.5 & 0.218 & -3.757 && 1.34~ &  -0.59  &   $_{-0.01}^{+0.01}$  & -0.57 & $_{-0.01}^{+0.01}$ & -0.27   & $_{-0.01}^{+0.01}$    \\
  10  &  053440.0-052604  &&    388~ & 235  &   152~ & 1118 & 0.88 && 742.7 & 0.420 & -6.062 && 0.99~ &  -0.61  &   $_{-0.31}^{+0.25}$  & -1.00 & $_{-...}^{+0.23}$ & \nodata & \nodata \\
  11  &  053440.8-052242  &&   5824~ & 123  &  5700~ &  691 & 0.88 && 693.2 & 0.392 & -4.415 && 1.53~ &  -0.40  &   $_{-0.01}^{+0.01}$  & -0.39 & $_{-0.01}^{+0.01}$ & -0.18   & $_{-0.02}^{+0.02}$    \\
  12  &  053440.8-052639  &&    374~ &  42  &   331~ &  228 & 0.39 && 702.0 & 0.397 & -5.297 && 1.79~ &  -0.15  &   $_{-0.06}^{+0.06}$  & -0.27 & $_{-0.07}^{+0.07}$ &  0.03   & $_{-0.09}^{+0.09}$    \\
  13  &  053440.9-052559  &&    325~ & 250  &    74~ &  970 & 0.89 && 746.2 & 0.422 & -6.380 && 1.11~ &  -1.00  &   $_{-...}^{+0.54}$  & -0.24 & $_{-0.29}^{+0.24}$ &  -1.00 & $_{-...}^{+0.89}$ \\
  14  &  053441.6-052357  &&    349~ & 136  &   212~ &  666 & 0.89 && 721.5 & 0.408 & -5.908 && 1.00~ &  -0.82  &   $_{-0.14}^{+0.14}$  & -0.80 & $_{-0.08}^{+0.09}$ & -0.69   & $_{-...}^{+0.65}$    \\
  15  &  053441.7-052652  &&   1758~ & 282  &  1475~ & 1103 & 0.88 && 698.5 & 0.395 & -5.034 && 1.29~ &  -0.45  &   $_{-0.04}^{+0.04}$  & -0.62 & $_{-0.03}^{+0.04}$ &  0.13   & $_{-0.08}^{+0.08}$    \\

\enddata

\tablenotetext{a}{The units for the counts-to-flux conversion
factor are $10^9$ ct cm$^2$ ph$^{-1}$.}

\tablenotetext{b}{The units for the log Incident Flux are ph
cm$^{-2}$ s$^{-1}$.}

\tablecomments{See \S \ref{photon_extraction_section} for
descriptions of the columns. The full table of 1616 sources is
provided in the electronic edition of the Journal.}
\end{deluxetable}

\clearpage
\newpage

\begin{deluxetable}{rrrrrr}
\centering \tabletypesize{\small} \tablewidth{0pt}
\tablecolumns{6}

\tablecaption{COUP Sources with Pileup
Analysis\label{pile-up_table}}

\tablehead{

\colhead{Seq} & \colhead{CXOONCJ} & \colhead{$r_{in}$} &
\colhead{$r_{out}$} &
\colhead{Net} & \colhead{PSF} \\

\colhead{\#} && \multicolumn{2}{c}{0.5\arcsec pix} &
\colhead{Counts}& \colhead{Frac}}

\startdata

 107 & 053455.9-052313 &  6.25 & 12.67 &   9004 & 0.08 \\
 123 & 053457.7-052351 &  5.00 & 11.95 &   5347 & 0.10 \\
 124 & 053457.7-052352 &  4.50 & 11.96 &   5074 & 0.13 \\
 245 & 053505.6-052519 &  3.50 & 11.73 &   2426 & 0.05 \\
 290 & 053507.4-052640 &  1.50 &  9.94 &   1895 & 0.51 \\
 394 & 053510.7-052344 &  2.50 &  9.14 &   4051 & 0.04 \\
 430 & 053511.5-052602 &  3.00 & 10.05 &   3621 & 0.06 \\
 450 & 053511.8-052149 &  4.00 &  9.95 &   5327 & 0.02 \\
 689 & 053515.2-052256 &  2.50 &  8.59 &   6931 & 0.03 \\
 724 & 053515.6-052256 &  4.00 &  8.54 &   3504 & 0.01 \\
 732 & 053515.7-052309 &  2.25 &  8.44 &  37523 & 0.04 \\
 745 & 053515.8-052314 &  2.25 &  8.39 &  21283 & 0.02 \\
 809\tablenotemark{a} & 053516.4-052322 &  5.25 &  8.43 &  63444 & 0.01 \\
 854 & 053517.0-052232 &  2.25 &  8.64 &   3829 & 0.05 \\
 855 & 053517.0-052334 &  3.00 &  8.75 &   3364 & 0.02 \\
 881 & 053517.3-052544 &  3.75 &  9.73 &   4521 & 0.02 \\
 891 & 053517.5-051740 &  6.00 & 16.56 &  12686 & 0.17 \\
 932 & 053517.9-052245 &  4.50 &  8.45 &   6648 & 0.01 \\
 965 & 053518.3-052237 &  3.50 &  8.56 &   5939 & 0.02 \\
1087 & 053520.1-052639 &  4.00 & 10.81 &   3792 & 0.02 \\
1090 & 053520.2-052056 &  3.00 & 10.31 &   4306 & 0.05 \\
1130 & 053521.0-052349 &  3.00 &  8.99 &   3843 & 0.03 \\
1232 & 053522.8-052457 &  4.25 &  9.66 &   3525 & 0.01 \\
1259 & 053523.6-052331 &  2.50 &  9.40 &   2183 & 0.04 \\

\enddata
\tablenotetext{a}{$= \theta^1$C Ori}

\end{deluxetable}

\clearpage
\newpage

\begin{deluxetable}{rrrcrrrrrrrrrrclc}
\centering \rotate \tabletypesize{\scriptsize} \tablewidth{0pt}
\tablecolumns{17}

\tablecaption{COUP Source X-ray Spectroscopy
\label{COUP_spe_table}}

\tablehead{

\colhead{Seq} & \colhead{CXOONCJ} & \colhead{$\log N_H$} &
\colhead{$\Delta \log N_H$} & \colhead{$kT_1$} & \colhead{$\Delta
kT_1$} & \colhead{$kT_2$} & \colhead{$\Delta kT_2$} &
\colhead{$\log EM_1$} & \colhead{$\pm$} & \colhead{$\log EM_2$} &
\colhead{$\pm$} & \colhead{$\chi^2_\nu$} & \colhead{dof} &
\colhead{Fit}&
\colhead{Model} & \colhead{Feature} \\

\colhead{\#} && \colhead{cm$^{-2}$} & \colhead{cm$^{-2}$} &
\colhead{keV} & \colhead{keV} & \colhead{keV} & \colhead{keV} &
\colhead{cm$^{-3}$} & \colhead{cm$^{-3}$} & \colhead{cm$^{-3}$} &
\colhead{cm$^{-3}$} &&
\colhead{Flag}& \colhead{Flag} & \colhead{Flag} \\

\colhead{(1)} & \colhead{(2)} & \colhead{(3)} & \colhead{(4)} &
\colhead{(5)} & \colhead{(6)} & \colhead{(7)} & \colhead{(8)} &
\colhead{(9)} & \colhead{(10)} & \colhead{(11)} & \colhead{(12)} &
\colhead{(13)} &  \colhead{(14)} & \colhead{(15)} & \colhead{(16)}
& \colhead{(17)}  }

\startdata

   1  & 053429.4-052337 & 20.67~ & 0.43 & 0.85 & 0.12 &  4.37 &  0.95 & 52.17~~ & 0.11 & 52.62~~ & 0.04 & 1.15 &  33 &\nodata& 20.1.2  & 000c0000 \\
   2  & 053429.5-052354 & 21.21~ & 0.07 & 0.79 & 0.03 &  3.29 &  0.35 & 53.13~~ & 0.06 & 53.16~~ & 0.04 & 1.15 &  23 &\nodata& 60.1.2  & 00000000 \\
   3  & 053433.9-052211 & 20.21~ & 2.31 & 1.14 & 0.27 &\nodata&\nodata& 51.28~~ & 0.25 &\nodata  &      & 0.70 &  24 &\nodata& 05.1.1  & 00h00000 \\
   4  & 053434.9-052507 & 23.39~ & 0.27 & 1.44 & 4.39 &\nodata&\nodata& 53.72~~ & 1.00 &\nodata  &      & 0.65 &  58 &\nodata& 05.1.1g & 0s000000 \\
   5  & 053438.2-052338 & 23.46~ & 0.21 & 9.52 &15.00 &\nodata&\nodata& 52.61~~ & 0.44 &\nodata  &      & 1.15 &  24 &\nodata& 05.1.1  & 0s000m00 \\
   6  & 053438.2-052423 & 21.12~ & 0.12 & 0.78 & 0.09 &  2.32 &  0.27 & 52.16~~ & 0.12 & 52.68~~ & 0.04 & 1.15 &  28 &\nodata& 60.1.2  & 000c0000 \\
   7  & 053439.7-052425 & 20.93~ & 0.04 & 0.83 & 0.01 &  2.30 &  0.08 & 53.58~~ & 0.02 & 53.94~~ & 0.01 & 2.06 & 173 &   m   & 60.1.2  & l00000w0 \\
   8  & 053439.8-052456 & 22.05~ & 0.05 &15.00 & 8.48 &\nodata&\nodata& 52.81~~ & 0.02 &\nodata  &      & 0.95 &  18 &\nodata& 60.1.1  & 00000000 \\
   9  & 053439.9-052642 & 21.37~ & 0.03 & 0.83 & 0.02 &  3.04 &  0.11 & 53.59~~ & 0.03 & 53.87~~ & 0.01 & 1.38 & 134 &   p   & 60.1.2g & 00000000 \\
  10  & 053440.0-052604 & 20.00~ & 1.21 & 0.74 & 0.13 &\nodata&\nodata& 51.42~~ & 0.21 &\nodata  &      & 0.77 &  44 &\nodata& 07.1.1  & 00h00000 \\
  11  & 053440.8-052242 & 21.69~ & 0.04 & 0.52 & 0.03 &  4.13 &  0.29 & 53.21~~ & 0.12 & 53.34~~ & 0.02 & 1.81 &  70 &   p   & 60.1.2  & l0000000 \\
  12  & 053440.8-052639 & 21.33~ & 0.18 &14.34 &11.68 &\nodata&\nodata& 52.44~~ & 0.05 &\nodata  &      & 0.68 &  32 &\nodata& 10.1.1  & 00000000 \\
  13  & 053440.9-052559 & 20.00~ & 2.12 &12.58 &15.00 &\nodata&\nodata& 51.51~~ & 0.17 &\nodata  &      & 1.02 &  55 &\nodata& 05.1.1  & 00h00p00 \\
  14  & 053441.6-052357 & 20.81~ & 0.14 & 0.69 & 6.24 &  9.23 & 15.00 & 51.64~~ & 1.00 & 51.19~~ & 0.21 & 0.75 &  25 &\nodata& 10.1.2  & 00hc0000 \\
  15  & 053441.7-052652 & 21.16~ & 0.44 & 0.66 & 0.08 & 15.00 & 15.00 & 52.35~~ & 0.16 & 52.59~~ & 0.03 & 1.60 &  21 &   m   & 60.1.2  & 00h00000 \\

\enddata
\tablecomments{See \S \ref{spectral_analysis_section} for
descriptions of the columns. The full table of 1616 sources is
provided in the electronic edition of the Journal.}
\end{deluxetable}

\clearpage
\newpage

\begin{deluxetable}{rrrrrrrrrrrrrrclc}
\centering \tabletypesize{\small} \tablewidth{0pt}
\tablecolumns{9}

\tablecaption{COUP Source X-ray Variability
\label{COUP_var_table}}

\tablehead{

\colhead{Seq} & \colhead{CXOONCJ} & \colhead{Gap} & \colhead{$\log
P_{KS}$} & \colhead{BBNum} & \colhead{BBMin} &
\colhead{$\Delta {\rm BBMin}$} & \colhead{BBMax} & \colhead{$\Delta {\rm BBMax}$}  \\

\colhead{\#} && \colhead{Flag} &&& \colhead{ct ks$^{-1}$} &
\colhead{ct ks$^{-1}$} & \colhead{ct ks$^{-1}$} &
\colhead{ct ks$^{-1}$} \\

\colhead{(1)} & \colhead{(2)} & \colhead{(3)} & \colhead{(4)} &
\colhead{(5)} & \colhead{(6)} & \colhead{(7)} & \colhead{(8)} &
\colhead{(9)}  }

\startdata

   1 &  053429.4-052337  & \nodata &  -4.00 &   7~~~ &   0.50~ &  0.02~ &   36.67~ &  1.52~ \\
   2 &  053429.5-052354  & \nodata &  -3.40 &   9~~~ &   1.18~ &  0.10~ &   64.52~ & 14.96~ \\
   7 &  053433.9-052211  & \nodata &  -0.95 &   1~~~ &   0.18~ &  0.01~ &    0.18~ &  0.01~ \\
   4 &  053434.9-052507  & \nodata &  -0.68 &   1~~~ &   0.39~ &  0.01~ &    0.39~ &  0.01~ \\
   7 &  053438.2-052338  & \nodata &  -0.19 &   1~~~ &   0.34~ &  0.01~ &    0.34~ &  0.01~ \\
   6 &  053438.2-052423  & \nodata &  -4.00 &   8~~~ &   0.91~ &  0.09~ &   10.51~ &  0.26~ \\
   7 &  053439.7-052425  & \nodata &  -4.00 &  36~~~ &  26.65~ &  2.60~ &  301.08~ & 55.54~ \\
   8 &  053439.8-052456  & \nodata &  -1.06 &   1~~~ &   1.61~ &  0.02~ &    1.61~ &  0.02~ \\
   9 &  053439.9-052642  & \nodata &  -4.00 & 105~~~ &   6.34~ &  1.47~ &  518.43~ & 45.68~ \\
  10 &  053440.0-052604  & \nodata &  -3.00 &   6~~~ &   0.33~ &  0.02~ &   21.46~ &  2.20~ \\
  11 &  053440.8-052242  & \nodata &  -4.00 &  19~~~ &   3.33~ &  0.90~ &   98.11~ &  2.35~ \\
  12 &  053440.8-052639  & \nodata &  -4.00 &   5~~~ &   0.29~ &  0.02~ &    2.08~ &  0.16~ \\
  13 &  053440.9-052559  & \nodata &  -1.01 &   1~~~ &   0.39~ &  0.01~ &    0.39~ &  0.01~ \\
  14 &  053441.6-052357  & \nodata &  -2.70 &   2~~~ &   0.38~ &  0.01~ &    1.20~ &  0.11~ \\
  15 &  053441.7-052652  & \nodata &  -4.00 &   8~~~ &   0.77~ &  0.23~ &    3.56~ &  0.23~ \\

\enddata
\tablecomments{See \S \ref{variability_analysis_section} for
descriptions of the columns. The full table of 1616 sources is
provided in the electronic edition of the Journal.}
\end{deluxetable}

\clearpage
\newpage

\begin{deluxetable}{rrrrrrr}
\centering  \tabletypesize{\small} \tablewidth{0pt}
\tablecolumns{7}

\tablecaption{COUP Source X-ray Luminosities
\label{COUP_lum_table}}

\tablehead{

\colhead{Seq} & \colhead{CXOONCJ} & \colhead{$\log L_s$} &
\colhead{$\log L_h$} & \colhead{$\log L_{h,c}$} & \colhead{$\log
L_t$} & \colhead{$\log L_{t,c}$} \\

\colhead{\#} && \colhead{erg s$^{-1}$} & \colhead{erg s$^{-1}$} &
\colhead{erg s$^{-1}$} & \colhead{erg s$^{-1}$} &
\colhead{erg s$^{-1}$} \\

\colhead{(1)} & \colhead{(2)} & \colhead{(3)} & \colhead{(4)} &
\colhead{(5)} & \colhead{(6)} & \colhead{(7)}  }

\startdata

   1 &   053429.4-052337  &   29.51 &  29.50 & 29.50 & 29.81 & 29.84 \\
   2 &   053429.5-052354  &   30.11 &  29.95 & 29.96 & 30.34 & 30.47 \\
   3 &   053433.9-052211  &   28.12 &  27.42 & 27.42 & 28.20 & 28.22 \\
   4 &   053434.9-052507  &$<$27.00 &  29.09 & 30.04 & 29.09 & 30.65 \\
   5 &   053438.2-052338  &$<$27.00 &  29.05 & 29.62 & 29.05 & 29.80 \\
   6 &   053438.2-052423  &   29.47 &  29.31 & 29.32 & 29.69 & 29.79 \\
   7 &   053439.7-052425  &   30.82 &  30.57 & 30.57 & 31.01 & 31.08 \\
   8 &   053439.8-052456  &   28.86 &  29.81 & 29.85 & 29.85 & 30.01 \\
   9 &   053439.9-052642  &   30.64 &  30.62 & 30.64 & 30.93 & 31.08 \\
  10 &   053440.0-052604  &   28.37 &  27.11 & 27.11 & 28.39 & 28.39 \\
  11 &   053440.8-052242  &   29.92 &  30.17 & 30.20 & 30.36 & 30.63 \\
  12 &   053440.8-052639  &   28.95 &  29.47 & 29.48 & 29.58 & 29.64 \\
  13 &   053440.9-052559  &   28.22 &  28.54 & 28.54 & 28.71 & 28.71 \\
  14 &   053441.6-052357  &   28.60 &  28.25 & 28.25 & 28.76 & 28.82 \\
  15 &   053441.7-052652  &   29.35 &  29.58 & 29.59 & 29.78 & 29.86 \\

\enddata
\tablecomments{See \S \ref{luminosities_section} for descriptions
of the columns. The full table of 1616 sources is provided in the
electronic edition of the Journal.}
\end{deluxetable}

\clearpage
\newpage

\begin{deluxetable}{rrrrrrcrrrrrrrrrrrr}
\centering \rotate \tabletypesize{\scriptsize} \tablewidth{0pt}
\tablecolumns{19}

\tablecaption{Probable Optical Counterparts of COUP Sources
\label{COUP_opt_table}}

\tablehead{

\colhead{Seq} & \colhead{CXOONCJ} & \colhead{ID} & \colhead{Off} &
\colhead{$V$} & \colhead{$I$} & \colhead{SpTy} & \colhead{$A_V$} &
\colhead{$\log T_{\rm eff}$} & \colhead{$\log L_{\rm bol}$} &
\colhead{$R$} & \colhead{$M$} & \colhead{$\log t$} &
\colhead{$\Delta (I-K)$} & \colhead{$EW({\rm Ca})$}&
\colhead{$<V>$} &
\colhead{$\Delta V$} & \colhead{$\pm$} & \colhead{$P$} \\

\colhead{\#} &&& \colhead{\arcsec} & \colhead{mag} & \colhead{mag}
&& \colhead{mag} & \colhead{$^\circ$K} & \colhead{L$_\odot$}&
\colhead{R$_\odot$} & \colhead{M$_\odot$} & \colhead{yr} &
\colhead{mag} & \colhead{$\AA$} & \colhead{mag} & \colhead{mag} &
\colhead{mag} & \colhead{day} \\

\colhead{(1)} & \colhead{(2)} & \colhead{(3)} & \colhead{(4)} &
\colhead{(5)} & \colhead{(6)} & \colhead{(7)} & \colhead{(8)} &
\colhead{(9)} & \colhead{(10)} & \colhead{(11)} & \colhead{(12)} &
\colhead{(13)} &  \colhead{(14)} & \colhead{(15)} & \colhead{(16)}
& \colhead{(17)} & \colhead{(18)} & \colhead{(19)}  }

\startdata

   1  & 053429.4-052337 &     8 & 0.39  &\nodata& 15.73 &  M5.5   & 0.00  & 3.483 & -0.98 &  1.18 &  0.14 &  6.16 &\nodata&   0.0 &\nodata&\nodata&\nodata&\nodata\\
   2  & 053429.5-052354 &     5 & 4.85  & 12.01 & 10.97 &  K1     & 0.29  & 3.708 &  0.64 &  2.69 &  1.98 &  6.58 &  0.01 &\nodata&\nodata&\nodata&\nodata&\nodata\\
   3  & 053433.9-052211 &    25 & 6.43  & 17.76 & 15.22 &  M4.5   & 0.00  & 3.500 & -0.88 &  1.21 &  0.17 &  5.95 &  0.52 &   0.0 &\nodata&\nodata&\nodata&\nodata\\
   4  & 053434.9-052507 &\nodata&\nodata&\nodata&\nodata& \nodata &\nodata&\nodata&\nodata&\nodata&\nodata&\nodata&\nodata&\nodata&\nodata&\nodata&\nodata&\nodata\\
   5  & 053438.2-052338 &\nodata&\nodata&\nodata&\nodata& \nodata &\nodata&\nodata&\nodata&\nodata&\nodata&\nodata&\nodata&\nodata&\nodata&\nodata&\nodata&\nodata\\
   6  & 053438.2-052423 &    40 & 0.36  & 17.37 & 14.48 &  M3.5   & 0.46  & 3.518 & -0.57 &  1.60 &  0.27 &  6.31 &  0.12 &   0.0 & 14.40 &  0.06 &  0.16 &  9.81 \\
   7  & 053439.7-052425 &    45 & 0.75  & 11.38 & 09.89 &  K1-K4  & 0.75  & 3.661 &  1.19 &  6.23 &  2.12 &  5.55 & -0.02 &\nodata&\nodata&\nodata&\nodata&\nodata\\
   8  & 053439.8-052456 &\nodata&\nodata&\nodata&\nodata& \nodata &\nodata&\nodata&\nodata&\nodata&\nodata&\nodata&\nodata&\nodata&\nodata&\nodata&\nodata&\nodata\\
   9  & 053439.9-052642 &    46 & 1.06  & 12.39 & 11.12 &  K0-K3  & 0.88  & 3.708 &  0.73 &  2.96 &  2.11 &  6.49 &  0.10 &\nodata&\nodata&\nodata&\nodata&\nodata\\
  10  & 053440.0-052604 &    48 & 1.02  & 20.10 & 15.50 &  M6     & 1.71  & 3.471 & -0.37 &  2.49 &  0.18 &  5.15 & -0.69 &   1.0 &\nodata&\nodata&\nodata&\nodata\\
  11  & 053440.8-052242 &    50 & 0.70  & 13.40 & 11.65 &  K1e-K7 & 0.42  & 3.602 &  0.44 &  3.46 &  0.69 &  5.74 &  1.37 & -14.6 &\nodata&\nodata&\nodata&\nodata\\
  12  & 053440.8-052639 &    51 & 1.36  & 16.82 & 14.45 &  M1.5   & 0.77  & 3.555 & -0.57 &  1.35 &  0.39 &  6.45 &  1.05 &   0.0 &\nodata&\nodata&\nodata&\nodata\\
  13  & 053440.9-052559 &    53 & 1.14  & 18.65 & 15.85 &  M0     & 2.58  & 3.580 & -0.71 &  1.02 &  0.55 &  6.93 &  1.13 &   1.5 &\nodata&\nodata&\nodata&\nodata\\
  14  & 053441.6-052357 &    54 & 0.76  & 18.85 & 15.27 &  M5     & 0.77  & 3.494 & -0.68 &  1.57 &  0.20 &  6.34 & -0.04 &   0.0 &\nodata&\nodata&\nodata&\nodata\\
  15  & 053441.7-052652 &    55 & 0.94  & 17.07 & 14.40 &  M3     & 0.31  & 3.526 & -0.60 &  1.48 &  0.29 &  6.33 &  0.36 &   1.0 &\nodata&\nodata&\nodata&\nodata\\

\enddata

\tablecomments{See \S \ref{stellar_counterparts_section} for
descriptions of the columns. The full table of 1616 sources is
provided in the electronic edition of the Journal.}
\end{deluxetable}

\clearpage
\newpage

\begin{deluxetable}{rrrrrrrrrrrrrrr}
\centering \rotate \tabletypesize{\scriptsize} \tablewidth{0pt}
\tablecolumns{15}

\tablecaption{Probable Near-Infrared Counterparts of COUP Sources
\label{COUP_nir_table}}

\tablehead{

\multicolumn{2}{c}{COUP} && \multicolumn{8}{c}{$JHK_s$ bands} &&
\multicolumn{3}{c}{L band} \\
\cline{1-2} \cline{4-11} \cline{13-15}

\colhead{Seq} & \colhead{CXOONCJ} && \colhead{Source} & \colhead{Off} &
\colhead{$J$} & \colhead{$H$} & \colhead{$K_s$} & \colhead{Src} &
\colhead{Phot} & \colhead{Con} && \colhead{Ref} & \colhead{Src} &
\colhead{$L$} \\

\colhead{\#} &&&& \colhead{\arcsec} & \colhead{mag} &
\colhead{mag} & \colhead{mag} & \colhead{Flag} & \colhead{Flag}& \colhead{Flag}&&
\colhead{Flag} & \colhead{\#} & \colhead{mag} \\

\colhead{(1)} & \colhead{(2)} && \colhead{(3)} & \colhead{(4)} &
\colhead{(5)} & \colhead{(6)} & \colhead{(7)} & \colhead{(8)} &
\colhead{(9)} & \colhead{(10)} && \colhead{(11)} & \colhead{(12)}
& \colhead{(13)} }

\startdata

   1  &  053429.4-052337 &&  05342945-0523374 &  0.04 &  13.99 &  13.37 &  13.09  & \nodata &  AAA  &  ccc   && \nodata&\nodata&\nodata \\
   2  &  053429.5-052354 &&  05342924-0523567 &  4.85 &  10.19 &   9.78 &   9.61  & \nodata &  AAA  &  000   && \nodata&\nodata&\nodata \\
   3  &  053433.9-052211 &&  05343357-0522087 &  6.25 &  13.25 &  12.54 &  12.17  & \nodata &  AAA  &  000   && \nodata&\nodata&\nodata \\
   4  &  053434.9-052507 &&     \nodata       &\nodata& \nodata& \nodata& \nodata & \nodata &\nodata&\nodata && \nodata&\nodata&\nodata \\
   5  &  053438.2-052338 &&     \nodata       &\nodata& \nodata& \nodata& \nodata & \nodata &\nodata&\nodata && \nodata&\nodata&\nodata \\
   6  &  053438.2-052423 &&  05343822-0524236 &  0.20 &  12.73 &  11.95 &  11.70  & \nodata &  AAA  &  0d0   && \nodata&\nodata&\nodata \\
   7  &  053439.7-052425 &&  05343976-0524254 &  0.48 &   8.85 &   8.10 &   7.95  & \nodata &  AAA  &  000   && \nodata&\nodata&\nodata \\
   8  &  053439.8-052456 &&     \nodata       &\nodata& \nodata& \nodata& \nodata & \nodata &\nodata&\nodata && \nodata&\nodata&\nodata \\
   9  &  053439.9-052642 &&  05343988-0526420 &  0.55 &  10.22 &   9.65 &   9.46  & \nodata &  AAA  &  000   && \nodata&\nodata&\nodata \\
  10  &  053440.0-052604 &&  05344000-0526040 &  0.45 &  13.33 &  12.70 &  12.34  & \nodata &  AAA  &  000   && \nodata&\nodata&\nodata \\
  11  &  053440.8-052242 &&  05344086-0522423 &  0.37 &  10.53 &   9.46 &   8.60  & \nodata &  AAA  &  000   && \nodata&\nodata&\nodata \\
  12  &  053440.8-052639 &&  05344081-0526387 &  0.77 &  12.68 &  11.67 &  11.14  & \nodata &  AAA  &  000   && \nodata&\nodata&\nodata \\
  13  &  053440.9-052559 &&  05344093-0526000 &  0.75 &  13.79 &  12.62 &  11.93  & \nodata &  AAA  &  000   && \nodata&\nodata&\nodata \\
  14  &  053441.6-052357 &&  05344162-0523574 &  0.30 &  13.28 &  12.57 &  12.28  & \nodata &  AAA  &  000   && \nodata&\nodata&\nodata \\
  15  &  053441.7-052652 &&  05344171-0526529 &  0.54 &  12.54 &  11.66 &  11.14  & \nodata &  AAA  &  000   && \nodata&\nodata&\nodata \\

\enddata

\tablecomments{See \S \ref{stellar_counterparts_section} for
descriptions of the columns. The full table of 1616 sources is
provided in the electronic edition of the Journal.}
\end{deluxetable}

\clearpage
\newpage

\begin{deluxetable}{rrrrrcrrcrrrrrrrr}
\centering \rotate \tabletypesize{\scriptsize} \tablewidth{0pt}
\tablecolumns{17} \tablecaption{Upper Limits of Undetected Stars
from Hillenbrand (1997)\label{COUP_lim opt_table}}

\tablehead{\colhead{H97-Srs} & \colhead{R.A.} & \colhead{Decl.} &
\colhead{LimCt} & \colhead{Exp} & \colhead{Conf} & \colhead{$V$} &
\colhead{$I$} & \colhead{SpTy} & \colhead{$A_V$} & \colhead{$\log
T_{\rm eff}$} & \colhead{$\log L_{\rm bol}$} & \colhead{$R$} &
\colhead{$M$} & \colhead{$\log t$} &
\colhead{$\Delta (I-K)$} & \colhead{$EW({\rm Ca})$}\\

\colhead{\#} &  \multicolumn{2}{c}{(degrees J2000)} &
\colhead{cts} & \colhead{ks} &\colhead{Flag} &  \colhead{mag} &
\colhead{mag} && \colhead{mag} & \colhead{$^\circ$K} &
\colhead{L$_\odot$}& \colhead{R$_\odot$} & \colhead{M$_\odot$} &
\colhead{yr} &
\colhead{mag} & \colhead{$\AA$}\\

\colhead{(1)} & \colhead{(2)} & \colhead{(3)} & \colhead{(4)} &
\colhead{(5)} & \colhead{(6)} & \colhead{(7)} & \colhead{(8)} &
\colhead{(9)} & \colhead{(10)} & \colhead{(11)} & \colhead{(12)} &
\colhead{(13)} &  \colhead{(14)} & \colhead{(15)} & \colhead{(16)}
& \colhead{(17)}}

\startdata
  9   & 83.623192 & -5.395426 &4017 & 396.2 & x       & 18.94   & 14.74 & K5      & 8.21    & 3.695   & 1.06    & 4.63    & 1.61    & 5.43    & \nodata  & 2.9   \\
  23  & 83.639984 & -5.385887 &  67 & 715.2 & \nodata & 13.08   & 12.00 & G7      & 0.82    & 3.740   & 0.38    & 1.71    & 1.50    & 6.83    & -0.10   & \nodata\\
  42  & 83.661568 & -5.356926 &  53 & 720.7 & \nodata & \nodata & 16.41 & G:      & \nodata & \nodata & \nodata & \nodata & \nodata & \nodata & \nodata & 4.7    \\
 3060 & 83.674896 & -5.358895 &  44 & 688.1 & x       & 18.60   & 16.27 & \nodata & \nodata & \nodata & \nodata & \nodata & \nodata & \nodata & \nodata & \nodata\\
 3073 & 83.677109 & -5.379485 &  46 & 746.7 & \nodata & \nodata & 16.60 & \nodata & \nodata & \nodata & \nodata & \nodata & \nodata & \nodata & \nodata & \nodata\\
  62  & 83.678688 & -5.421204 &  55 & 749.9 & \nodata & 17.64   & 14.59 & early-K & 5.26    & 3.695   & 0.40    & 2.17    & 1.38    & 6.21    & -0.10   & 1.5    \\
  67  & 83.682404 & -5.424037 &  53 & 752.6 & \nodata & \nodata & 16.62 & \nodata & \nodata & \nodata & \nodata & \nodata & \nodata & \nodata & \nodata & \nodata\\
  86  & 83.694984 & -5.434676 &  45 & 760.4 & \nodata & 18.64   & 15.83 & \nodata & \nodata & \nodata & \nodata & \nodata & \nodata & \nodata & \nodata & \nodata\\
  85  & 83.695068 & -5.358148 & 130 & 756.8 & x       & 17.39   & 14.78 & M2.5    & 0.59    & 3.535   & -0.71   & 1.25    & 0.26    & 6.15    & 0.14    & 4.1    \\
 108  & 83.708405 & -5.312427 &  66 & 454.9 & x       & 10.19   &  9.29 & A2-A7   & 2.07    & 3.940   & 2.01    & 4.46    & 3.59    & \nodata & 0.63    & \nodata\\
 141  & 83.729233 & -5.449654 & 208 & 667.4 & x       & 18.27   & 15.43 & G:      & \nodata & \nodata & \nodata & \nodata & \nodata & \nodata & \nodata & 1.3    \\
 179  & 83.741188 & -5.478287 & 237 & 731.5 & x       & \nodata & 16.65 & M5.5    & 0.00    & 3.483   & -1.34   & 0.77    & 0.09    & 6.32    & \nodata & 0.0    \\
 5084 & 83.754486 & -5.340973 &  24 & 789.3 & \nodata & \nodata & 16.99 & \nodata & \nodata & \nodata & \nodata & \nodata & \nodata & \nodata & \nodata & \nodata\\
 214  & 83.757027 & -5.443426 &  23 & 702.6 & \nodata & \nodata & 16.75 & \nodata & \nodata & \nodata & \nodata & \nodata & \nodata & \nodata & \nodata & \nodata\\
 219  & 83.758736 & -5.306343 &  35 & 765.1 & \nodata & 19.82   & 15.52 & M5      & 2.62    & 3.494   & -0.33   & 2.35    & 0.15    & 4.43    & 0.12    & 0.3    \\
\enddata

\tablecomments{The full table of 201 sources is provided in the
electronic edition of the Journal.}
\end{deluxetable}

\clearpage
\newpage

\begin{deluxetable}{rrrcrrrrr}
\centering  \tabletypesize{\scriptsize} \tablewidth{0pt}
\tablecolumns{9}

\tablecaption{Upper Limits of Undetected Stars from the Two Micron
All Sky Survey \label{COUP_lim 2m_table}}

\tablehead{

\colhead{2M-Src} & \colhead{LimCt} & \colhead{Exp} &
\colhead{Conf} & \colhead{$J$} & \colhead{$H$} & \colhead{$K_s$}
& \colhead{Phot} & \colhead{Con}\\

\colhead{} & \colhead{cts} & \colhead{ks}& \colhead{Flag} &
\colhead{mag} & \colhead{mag} &
\colhead{mag} & \colhead{Flag} & \colhead{Flag}\\

\colhead{(1)} & \colhead{(2)} & \colhead{(3)} & \colhead{(4)} &
\colhead{(5)} & \colhead{(6)} & \colhead{(7)} & \colhead{(8)} &
\colhead{(9)}}

\startdata
05342954-0523437 &  3891 & 383.9  &   x     & 11.90 & 10.70 & 10.11 & AAA & 000\\
05343297-0522142 &   475 & 126.4  &\nodata  & 16.63 & 15.30 & 14.60 & CUU & c00\\
05343359-0523099 &    66 & 715.2  &\nodata  & 11.28 & 10.78 & 10.72 & AAA & 000\\
05343419-0524593 &   615 & 490.2  &\nodata  & 18.65 & 16.20 & 15.13 & UCB & 000\\
05343508-0521498 &  1044 & 200.6  &\nodata  & 16.95 & 16.34 & 15.51 & UUC & 000\\
05343646-0521458 &   454 & 485.4  &\nodata  & 16.77 & 15.87 & 15.16 & BBB & 000\\
05343725-0525101 &    59 & 729.6  &\nodata  & 16.88 & 15.60 & 15.28 & CBB & 000\\
05343824-0524024 &   150 & 733.4  &\nodata  & 15.96 & 14.70 & 14.23 & AAA & 000\\
05343850-0523255 &   132 & 713.1  &   x     & 16.07 & 14.53 & 14.06 & AAA & 000\\
05343873-0521250 &    53 & 719.4  &\nodata  & 14.10 & 13.05 & 12.67 & AAA & 000\\
05344062-0523150 &    50 & 720.5  &\nodata  & 16.14 & 15.17 & 14.84 & AAA & 000\\
05344065-0526526 &   657 & 445.8  &   x     & 16.73 & 16.26 & 15.44 & BCB & ccc\\
05344135-0524442 &    58 & 744.9  &\nodata  & 16.87 & 15.91 & 15.50 & CCB & ccc\\
05344196-0521321 &    44 & 687.5  &\nodata  & 14.16 & 13.48 & 13.12 & AAA & 000\\
05344224-0522325 &    46 & 745.1  &\nodata  & 16.50 & 15.14 & 14.64 & BAA & c00\\
\enddata

\tablecomments{The full table of 1145 sources is provided in the
electronic edition of the Journal.  Note that some of these
sources are bright spots in the diffuse nebular emission rather
than stellar photospheres.}

\end{deluxetable}

\end{document}